\documentclass[%
 reprint,
 onecolumn,
 groupedaddress,
 unsortedaddress,
 nofootinbib,
 amsmath,amssymb,
 aps,
 prd,
]{revtex4-1}

\usepackage{graphicx}
\usepackage{dcolumn}
\usepackage{bm}
\usepackage[normalem]{ulem}
\usepackage{times}
\usepackage{color}
\usepackage{hyperref}
\usepackage{verbatim} 
\usepackage{slashed}
\begin{document}

\preprint{APS/123-QED}

\title{New K\"ahler invariant Fayet-Iliopoulos terms in supergravity and cosmological applications}

\author{I.~Antoniadis$^{\dag,\ddag}$, F.~Rondeau$^{\dag}$}
\email{antoniad@lpthe.jussieu.fr, francois.rondeau@lpthe.jussieu.fr}
\affiliation{$\dag$ Laboratoire de Physique Th\'eorique et Hautes Energies - LPTHE\\ Sorbonne Universit\'e, CNRS, 4 Place Jussieu, 75005 Paris, France.\\$^\ddag$ Albert Einstein Center for Fundamental Physics, Institute for Theoretical Physics, University of Bern, Sidlerstrasse 5, CH-3012 Bern, Switzerland}

\begin{abstract}

Recently, a new type of constant Fayet-Iliopoulos (FI) terms was introduced in $\mathcal{N}=1$ supergravity, which do not require the gauging of the $R$-symmetry. We revisit and generalise these constructions, building a new class of K\"ahler invariant FI terms parametrised by a function of the gravitino mass as functional of the chiral superfields, which is then used to describe new models of inflation. They are based on a no-scale supergravity model of the inflaton chiral multiplet, supplemented by an abelian vector multiplet with the new FI-term. We show that the inflaton potential is compatible with the CMB observational data, with a vacuum energy at the minimum that can be tuned to a tiny positive value. Finally, the axionic shift  symmetry can  be gauged by  the $U(1)$ which becomes massive. These models offer a mechanism for fixing the gravitino mass in no-scale supergravities, that corresponds to a flat direction of the scalar potential in the absence of the new FI-term; its origin in string theory is an interesting open problem.

\end{abstract}

\pacs{Valid PACS appear here}
\maketitle


\section{Introduction}
\label{sec:introduction}

The simplest extension of pure $\mathcal{N}=1$ supergravity in flat spacetime is the anti de Sitter (AdS) supergravity, where a negative cosmological constant $\Lambda$ is included \cite{Van_Proeyen}. In order to preserve local supersymmetry, a gravitino effective mass term has to be added, linked to $\Lambda$ through $\Lambda=-3m_{3/2}^2$, which describes a massless gravitino in AdS spacetime. It is simply obtained by considering a constant superpotential $W=m_{3/2}$. An arbitrary cosmological constant cannot be introduced without breaking explicitly supersymmetry, or considering non-linear realisation~\cite{Antoniadis:2014oya}. In the presence of an abelian vector multiplet a constant Fayet-Iliopoulos (FI) term can be introduced only if the $U(1)$ gauges the R-symmetry, in which case a constant superpotential is forbidden, leading to a de Sitter (dS) supergravity describing a massive gravitino through curvature effects~\cite{Freedman_model_1,Freedman_model_2}.

Recently, a new type of constant FI-term was introduced which does not require the gauging of the $R$-symmetry~\cite{new_FI_1}. It assumes that the D-auxiliary component of the $U(1)$ vector multiplet has a non-vanishing vacuum expectation value (VEV) breaking spontaneously supersymmetry, in which case it can be expanded as $D$ + fermion terms of higher dimensions. In the unitary gauge where the gravitino absorbs the $U(1)$ gaugino and becomes massive, the  fermion terms vanish and the new FI-term amounts adding a positive contribution to the cosmological constant of the AdS supergravity, since a constant superpotential is now allowed as the $U(1)$ does not gauge the R-symmetry. In the presence of matter, the construction of~\cite{new_FI_1} leads to a scalar potential but breaks K\"ahler invariance. On the other hand, the new and standard FI-terms can coexist in the case of gauge R-symmetry, providing interesting models of D-term inflation~\cite{Antoniadis:2018cpq}. An alternative construction was made in~\cite{new_FI_2} preserving K\"ahler invariance and leading to a constant FI-term in the presence of matter, that generates a constant uplift of the vacuum energy. More recently, such FI-terms were written in $\mathcal{N}=2$ supergravity exhibiting a much richer structure~\cite{Antoniadis:2019hbu}.

In this work, we generalise the above constructions in $\mathcal{N}=1$ supergravity, preserving the K\"ahler invariance and keeping the form of the bosonic action to be linear in $D$ up to a field dependent coefficient. We show that the most general FI-term is characterised by an arbitrary function of the gravitino mass, taken as a functional of the chiral superfields. We then study applications to cosmology, building new models of inflation compatible with CMB observations and possessing a dS vacuum with tuneable (tiny) energy. We specialise to no-scale models~\cite{Cremmer:1983bf} of one chiral multiplet containing the inflaton, supplemented by a $U(1)$ gauge symmetry with the new FI-term. Moreover, we choose for the latter a simple characteristic function of the gravitino mass which is a single power and an additive constant, thus depending on three parameters. We show that there is a region in the parameter space where the resulting scalar potential possesses an inflationary plateau describing successfully the cosmological observations with the inflaton rolling down to a minimum with tuneable vacuum energy, where  the gravitino mass and the supersymmetry breaking scale are fixed in terms of the
parameters of the model.

This paper is organised as follows. In Section \ref{sect:review}, we review the recent construction of the new FI-term in $\mathcal{N}=1$ supergravity without gauging the $R$-symmetry, and its generalisation to a K\"ahler invariant FI Lagrangian leading to a positive constant uplift of the scalar potential in the presence of arbitrary matter chiral multiplets. In Section \ref{sect:new_FI}, we propose the most general modification of this construction that preserves K\"ahler invariance and is characterised by an arbitrary function of the gravitino mass as functional of chiral multiplets. We then study the consequences of such terms on inflation and supersymmetry breaking in a de Sitter vacuum with tuneable energy in Section \ref{sect:inflation}, for the case of two no-scale models and for a simple choice of the functional dependence of the new FI D-term. Finally in Section \ref{sect:gauging_shift_symmetry}, we discuss the gauging of the shift symmetry that gets rid of the massless particles in the spectrum without altering the inflationary predictions. Moreover, inspired by the low-energy limit of the heterotic string, we identify the inflaton with the string dilaton and gauge the perturbative axionic symmetry by the Green-Schwarz anomaly cancellation mechanism. These models provide new examples of inflation by supersymmetry breaking~\cite{Antoniadis:2017gjr}, where the inflaton belongs to the same multiplet with the Goldstino~\cite{AlvarezGaume:2010rt}, without gauging the R-symmetry. Our conclusions are presented in  Section \ref{sect:conclusion}. There are also two appendices containing a summary of the conformal supergravity multiplets calculus (Appendix \ref{sect:append_multiplet_calculus}) and details of the computation of the fermion masses in our models (Appendix \ref{sect:append_fermion_masses}).\\

Throughout this work, we will use natural units $\hbar=c\equiv 1$. The reduced Planck mass $\kappa^{-1}=(8\pi G)^{-1/2}=2.4\times 10^{18}{\rm GeV}$ is set equal to one, and numerical values are given in these units. We adopt the metric convention $(-,+,+,+)$.

\section{Fayet-Iliopoulos term without gauged $R$-symmetry: a review}
\label{sect:review}
In $\mathcal{N}=1$ supergravity, a new Fayet-Iliopoulos term associated to a non-gauged $R$ symmetry has first been introduced in \cite{new_FI_1}. In the superconformal formalism, denoting $S_0=(s_0,P_L\Omega_0,F_0)$ and $\bar{S}_0=(\bar{s}_0,P_R\Omega_0,\bar{F}_0)$ the chiral and anti-chiral compensator superfields, with (Weyl, Chiral) weights $(1,1)$ and $(1,-1)$ respectively, this new FI Lagrangian reads~\cite{new_FI_1}:
\begin{equation}\label{eq:FI_lagrangian_1}
\mathcal{L}_{FI}=-\xi\left[S_0\bar{S}_0\frac{\mathcal{W}^2\bar{\mathcal{W}^2}}{T(\bar{\mathcal{W}^2})\bar{T}(\mathcal{W}^2)}(V)_D \right]_D,
\end{equation}
where $\xi$ is a constant parameter, $(V)_D$ is a real linear multiplet defined by $(V)_D=(D,\slashed{\mathcal{D}}\lambda,0,\mathcal{D}^b\hat{F}_{ab},-\slashed{\mathcal{D}}\slashed{\mathcal{D}}\lambda,-\Box^C D)$, whose lowest component $D$ is the real auxiliary field of the vector superfield $V$, the latter having (anti)-chiral field strength ($\bar{\mathcal{W}}$) $\mathcal{W}$
 given by 
\begin{equation}\label{eq:field_strength_1}
\mathcal{W}^2=\frac{\bar{\lambda}P_L\lambda}{S_0^2},~~~\bar{\mathcal{W}^2}=\frac{\bar{\lambda} P_R\lambda}{\bar{S}_0^2}\,,
\end{equation}
so that $(V)_D$ is given by the super-covariant derivative of $\mathcal{W}$.
The chiral projection operator $T$ acting on an anti-chiral multiplet $\bar{X}=(\bar{X},P_R\Omega,\bar{F})$ of weights $(1,-1)$ gives a chiral multiplet of weights $(2,2)$ defined as $T(\bar{X})=(\bar{F},\slashed{\mathcal{D}}P_R\Omega,\Box^C\bar{X})$. $\bar{\lambda}P_L\lambda$ has weights $(3,3)$ and reads, in components form:
\begin{equation}\label{eq:lambda^2}
\bar{\lambda}P_L\lambda=\left(\bar{\lambda}P_L\lambda;\sqrt 2P_L\left(-\frac{1}{2}\gamma\cdot\hat{F}+iD\right)\lambda;2\bar{\lambda}P_L\slashed{\mathcal{D}}\lambda+\hat{F}^-\cdot\hat{F}^--D^2\right),    
\end{equation}
with the covariant field strength $\hat{F}_{ab}$ and the self-dual and anti self-dual tensors $\hat{F}_{ab}^{\pm}$ given by
\begin{eqnarray}
\hat{F}_{ab}&=&e_a^{\mu}e_b^{\nu}(2\partial_{[\mu}A_{\nu]}+\bar{\psi}_{[\mu}\gamma_{\nu]}\lambda),\\
\hat{F}_{ab}^{\pm}&=&\frac{1}{2}(\hat{F}_{ab}\pm\tilde{\hat{F}}_{ab}).
\end{eqnarray}
The dual field strength is $\tilde{\hat{F}}_{ab}=-\frac{i}{2}\epsilon_{abcd}\hat{F}^{cd}$, while the covariant derivative $\mathcal{D}_{\mu}\lambda$ is defined by
\begin{equation}
\mathcal{D}_{\mu}\lambda=\left(\partial_{\mu}-\frac{3}{2}b_{\mu}+\frac{1}{4}w_{\mu}^{ab}\gamma_{ab}-\frac{3}{2}i\gamma_*\mathcal{A}_{\mu}\right)\lambda-\left(\frac{1}{4}\gamma^{ab}\hat{F}_{ab}+\frac{1}{2}i\gamma_*D\right)\psi_{\mu}.
\end{equation}
$\psi_{\mu}$ is the gravitino, the fields $b_{\mu}$, $w_{\mu}^{ab}$ and $\mathcal{A}_{\mu}$ are the gauge fields corresponding to dilatations, Lorentz transformations and $T_R$ symmetry of the conformal algebra respectively, and $\gamma_*\equiv i\gamma_0\gamma_1\gamma_2\gamma_3$.

Considering first pure supergravity coupled to an abelian $U(1)$ gauge multiplet plus the FI term \eqref{eq:FI_lagrangian_1}, the full Lagrangian reads:
\begin{equation}\label{eq:AdS_supergravity_plus_new_FI}
\mathcal{L}=-3\left[S_0\bar{S}_0\right]_D+\left[S_0^3W_0\right]_F-\frac{1}{4g^2}\left[\bar{\lambda}P_L\lambda\right]_F+\mathcal{L}_{FI}.
\end{equation}
Supersymmetry is broken via a non-vanishing VEV of the D-auxiliary component of the vector multiplet driven by the linear term in D, with the Goldstino being the $U(1)$ gaugino.
In component form, after having gauge fixed the compensator through $S_0=1$, integrated the auxiliary fields, and in the unitary gauge where the Goldstino vanishes, one gets \cite{new_FI_1}:
\begin{equation}
e^{-1}\mathcal{L}=\frac{1}{2}\left(R-\bar{\psi}_{\mu}\gamma^{\mu\nu\rho}D_{\nu}\psi_{\rho}+m_{3/2}\bar{\psi}_{\mu}\gamma^{\mu\nu}\psi_{\nu}\right)-\frac{1}{4g^2}F^{\mu\nu}F_{\mu\nu}-\left(-3m_{3/2}^2+\frac{1}{2}\xi^2\right)
\end{equation}
with $m_{3/2}=W_0$, a constant superpotential.
Therefore, in the absence of chiral matter superfields, any $\xi\neq 0$ uplifts the vacuum energy by a constant term $\mathcal{V}_{FI}={\xi^2}/{2}$ and breaks supersymmetry. One can then tune $\xi$ to get a de Sitter vacuum configuration, matching with the observational data. For instance, $\xi=\sqrt{6}m_{3/2}$ gives a massive gravitino in flat Minkowski spacetime with spontaneously broken supersymmetry. 

Introducing chiral matter multiplets $X^i$ in the previous model, the Lagrangian is now given by:
\begin{equation}\label{eq:AdS_supergravity_plus_new_FI_plus_chiral_matter}
\mathcal{L}=-3\left[S_0\bar{S}_0e^{-K(X^i,\bar{X^i})/3}\right]_D+\left[S_0^3W(X^i)\right]_F-\frac{1}{4g^2}\left[\bar{\lambda}P_L\lambda\right]_F+\mathcal{L}_{FI}.
\end{equation}
In component form, after having gauge fixed the compensator through $S_0=e^{K/6}$ and integrated the auxiliary fields, the bosonic part of the previous Lagrangian reads \cite{new_FI_1}:
\begin{equation}
\left.e^{-1}\mathcal{L}\right|_{\text{bos}}=\frac{1}{2}R-\frac{1}{4g^2}F^{\mu\nu}F_{\mu\nu}-G_{i\bar j}\partial X^i\cdot\partial \bar{X}^{\bar j}-\left(e^K(|\nabla_iW|^2-3|W|^2)+\frac{\xi^2g^2}{2}e^{2/3K}\right).
\end{equation}
Therefore, when matter fields are coupled, the scalar potential contribution from \eqref{eq:FI_lagrangian_1} becomes field dependent, $\mathcal{V}_{FI}=\frac{\xi^2g^2}{2}e^{2K/3}$, and no longer K\"ahler invariant, which basically comes from the fact that the FI Lagrangian \eqref{eq:FI_lagrangian_1} is not itself K\"ahler invariant. To remedy this, a generalized K\"ahler invariant FI term has been built in \cite{new_FI_2}. From the generic K\"ahler transformations for a K\"ahler potential $K(X,\bar X)$, a superpotential $W(X)$ and the compensator $S_0$,
\begin{eqnarray}\label{eq:Kahler_transfo}
K(X,\bar X)&\rightarrow&K(X,\bar X)+J(X)+\bar J(\bar X),\nonumber\\
W(X)&\rightarrow&W(X)e^{-J(X)},\\
S_0&\rightarrow&S_0e^{J(X)/3},\nonumber
\end{eqnarray}
this new construction is based on the modification of the FI term \eqref{eq:FI_lagrangian_1} by introducing in it the K\"ahler potential according to
\begin{equation}\label{eq:FI_lagrangian_2}
\mathcal{L}_{FI}=-\xi\left[(S_0\bar{S}_0e^{-K/3})^{-3}\frac{(\bar{\lambda}P_L\lambda)(\bar{\lambda} P_R\lambda)}{T(\bar{\mathcal{W}'^2})\bar{T}(\mathcal{W}'^2)}(V)_D \right]_D.
\end{equation}
The modified and henceforth K\"ahler invariant gauge field strengths are given by\footnote{Note that these superfields are no longer chiral, contrary to the previous ones of Eq.~\eqref{eq:field_strength_1}.}
\begin{equation}\label{eq:field_strength_2}
\mathcal{W}'^2=\frac{\bar{\lambda}P_L\lambda}{(S_0\bar{S_0}e^{-K/3})^2},~~~\bar{\mathcal{W}}'^2=\frac{\bar{\lambda} P_R\lambda}{(S_0\bar{S_0}e^{-K/3})^2}.
\end{equation}
The new bosonic contribution to the scalar potential arising from this new term reads $\mathcal{V}_{FI}=\frac{\xi^2g^2}{2}$, which is constant whether matter fields are included or not. The aim of this letter is to generalise the work carried out in \cite{new_FI_1} and \cite{new_FI_2} by building the most general extended FI terms whose bosonic component is linear in the auxiliary field D, up to a general field dependent coefficient, while preserving K\"ahler invariance at the same time. 

\section{A set of K\"ahler invariant Fayet-Iliopoulos terms}
\label{sect:new_FI}
The starting point of the approach followed in this work is to modify the field strengths \eqref{eq:field_strength_1} by introducing the superpotential $W$ in order to make them K\"ahler invariant. This can be done in the following way\footnote{Besides K\"ahler invariance, the new gauge field strengths \eqref{eq:field_strength_3} are again chiral like those of Eq.~\eqref{eq:field_strength_1}.}:
\begin{equation}\label{eq:field_strength_3}
\mathcal{W}^2=\frac{\bar{\lambda}P_L\lambda}{{S}_{0}^2 W(X)^{\frac{2}{3}}},~~~\bar{\mathcal{W}^2}=\frac{\bar{\lambda} P_R\lambda}{\bar{S}_{0}^2\bar{W}(\bar{X)}^{\frac{2}{3}}},
\end{equation}
where the $\frac{2}{3}$ exponent of $W(X)$ is uniquely fixed by the K\"ahler transformations \eqref{eq:Kahler_transfo} to get K\"ahler invariant $\mathcal{W}^2$ and $\bar{\mathcal{W}}^2$. The superpotential $W$ has vanishing Weyl and Chiral weights and is assumed to have a non-vanishing VEV. Therefore, $\mathcal{W}^2$ and $\bar{\mathcal{W}}^2$ have the same (Weyl, Chiral) weights $(1,1)$ and $(1,-1)$ as those of \eqref{eq:field_strength_1}, and one can thus still apply the (anti-) chiral projection operators ($\bar{T}$) $T$. The resulting multiplets $T(\bar{\mathcal{W}}^2)$ and $\bar{T}(\mathcal{W}^2)$ then carry weights $(2,2)$ and $(2,-2)$. The operation $[~]_D$ has to act on a multiplet of weights $(2,0)$. $(V)_D$ having already weights $(2,0)$, we need to multiply it with a multiplet with vanishing weights, which can be chosen $(S_0\bar{S}_0)^{-1}\frac{(\bar{\lambda}P_L\lambda)(\bar{\lambda} P_R\lambda)}{T(\bar{\mathcal{W}}^2)\bar{T}(\mathcal{W}^2)}$ as it can be easily checked knowing the weights $(1,1)$ and $(3,3)$ of $S_0$ and $\bar{\lambda}P_L\lambda$, respectively. The K\"ahler potential $K$ and the superpotential $W$ having vanishing weights, we can include them for free in the previous combination in the following form:
\begin{equation}
\mathcal{R}(V)_D\equiv(S_0\bar{S}_0)^{-1}e^{nK}W^{\alpha}\bar{W}^{\alpha}\frac{(\bar{\lambda}P_L\lambda)(\bar{\lambda} P_R\lambda)}{T(\bar{\mathcal{W}^2})\bar{T}(\mathcal{W}^2)}(V)_D,
\end{equation}
where at this point the parameters $n$ and $\alpha$ are arbitrary. As it can be seen from \eqref{eq:Kahler_transfo}, the quantity $\mathcal{R}(V)_D$ is K\"ahler invariant provided that $n$ and $\alpha$ are related by
\begin{equation}\label{eq:Kahler condition}
n=\frac{1}{3}+\alpha.
\end{equation}

Therefore, the most general K\"ahler invariant FI term involving both the K\"ahler potential and the superpotential is in fact a set of Lagrangians labelled by one free parameter $\alpha$ according to:
\begin{equation}\label{eq:FI_lagrangian_3}
\mathcal{L}_{FI}^{(\alpha)}=-\xi_\alpha\left[(S_0\bar{S}_0)^{-1}e^{(\frac{1}{3}+\alpha)K}W^{\alpha}\bar{W}^{\alpha}\frac{(\bar{\lambda}P_L\lambda)(\bar{\lambda} P_R\lambda)}{T(\bar{\mathcal{W}^2})\bar{T}(\mathcal{W}^2)}(V)_D \right]_D.
\end{equation}
We now add a series of terms \eqref{eq:FI_lagrangian_3} in $\mathcal{N}=1$ supergravity coupled to the $U(1)$ gauge multiplet (whose gauge kinetic function is chosen to be one for simplicity), plus a set of matter chiral multiplets denoted generically $\{X\}$. Before gauge fixing the superconformal generators, the lagrangian for this model reads:
\begin{equation}\label{eq:full_lagrangian}
\mathcal{L}=-3\left[S_0\bar{S}_0 e^{-\frac{\mathcal{K}(X,\bar X)}{3}}\right]_D+\left[S_0^3W(X)\right]_F-\frac{1}{4g^2}\left[\bar{\lambda}P_L\lambda\right]_F+\underset{i}{\sum} \mathcal{L}_{FI}^{(\alpha_i)},
\end{equation}
where the sum is running for now over an arbitrary set of parameters $\alpha_i$. 

We are interested in the contribution of Eq.~\eqref{eq:FI_lagrangian_3} to the scalar potential, and in particular we would like to check that K\"ahler invariance is preserved. For simplicity and in order to highlight the cosmological applications, we focus on the bosonic sector. The contribution to the fermion masses arising from these new FI terms is studied in Appendix~\ref{sect:append_fermion_masses}. Putting all fermions to zero for now, the remaining components of the chiral multiplet $\bar{\lambda}P_L\lambda$ given in Eq.~\eqref{eq:lambda^2} are:
\begin{equation}
\bar{\lambda}P_L\lambda=\left(0,~0,~\hat{F}^-\cdot\hat{F}^--D^2\right).    
\end{equation}
With the chiral and anti-chiral compensators $S_0=(s_0,P_L\Omega_0,F_0)$ and $\bar{S}_0=(\bar{s}_0,P_R\Omega_0,\bar{F}_0)$, the bosonic components of the composite chiral and anti-chiral multiplets $S_0^{-1}\bar{\lambda}P_L\lambda$ and $\bar{S}_0^{-1}\bar{\lambda} P_R\lambda$ are given by\footnote{The three and seven-components notation for chiral multiplets, as well as the main calculus rules used in this paper, are described in Appendix \ref{sect:append_multiplet_calculus}.}:
\begin{eqnarray}
\label{eq:1}
S_0^{-1}\bar{\lambda}P_L\lambda&=&(0,~0,~s_0^{-1}(\hat{F}^-\cdot\hat{F}^--D^2))=(0,~0,~-2s_0^{-1}(\hat{F}^-\cdot\hat{F}^--D^2),~0,~0,~0,~0),\\
\label{eq:2}
\bar{S}_0^{-1}\bar{\lambda} P_R\lambda&=&(0,~0,~\bar{s}_0^{-1}(\hat{F}^+\cdot\hat{F}^+-D^2))=(0,~0,~0,~-2\bar{s}_0^{-1}(\hat{F}^+\cdot\hat{F}^+-D^2)~0,~0,~0).
\end{eqnarray}
Similarly, the bosonic components of the chiral superfield $\mathcal{W}^2$ given in Eq.~\eqref{eq:field_strength_3} are:
\begin{equation}
\mathcal{W}^2=\left(0,~0,~s_0^{-2}W^{-\frac{2}{3}}(\hat{F}^-\cdot\hat{F}^--D^2)\right),
\end{equation}
from which we deduce the components of the anti-chiral superfield $\bar{T}(\mathcal{W}^2)$:
\begin{equation}
\bar{T}(\mathcal{W}^2)=\left(s_0^{-2}W^{-\frac{2}{3}}(\hat{F}^-\cdot\hat{F}^--D^2),~0,~0\right).
\end{equation}
The product of the multiplets \eqref{eq:1} and \eqref{eq:2} already being a $\theta^2\bar{\theta}^2$ term, only the lowest components of the remaining quantities entering in the Lagrangian \eqref{eq:FI_lagrangian_3} will contribute to the bosonic sector. We can therefore rewrite it as:
\begin{equation}\label{eq:bosonic_FI_lagrangian}
\mathcal{L}_{FI}^{(\alpha_i)}=-\xi_i\frac{e^{(\frac{1}{3}+\alpha_i)K}W^{\alpha_i}\bar{W}^{\alpha_i}D}{(s_0\bar{s}_0)^{-2}(W\bar{W})^{-\frac{2}{3}}(\hat{F}^-\cdot \hat{F}^--D^2)(\hat{F}^+\cdot \hat{F}^+-D^2)}\left[\mathcal{R}\right]_D,
\end{equation}
with the real multiplet $\mathcal{R}$ defined as: 
\begin{equation}
\mathcal{R}\equiv(S_0^{-1}\bar{\lambda}P_L\lambda)(\bar{S}_0^{-1}\bar{\lambda} P_R\lambda).
\end{equation}

Looking at the seven-components notation \eqref{eq:1} and \eqref{eq:2} for the multiplets $S_0^{-1}\bar{\lambda}P_L\lambda$ and $\bar{S}_0^{-1}\bar{\lambda} P_R\lambda$ and the multiplication law \eqref{eq:multiplet_composition_laws}, we see that the only non-vanishing \emph{bosonic} term of $\mathcal{R}$ arises from $\frac{1}{2}f_{ij}\mathcal{K}^i\mathcal{H}^j$ in its D-component. More precisely, it reads $(\mathcal{R})_D=\frac{1}{2}f_{ij}\mathcal{K}^i\mathcal{H}^j=\frac{1}{2}f_{21}\mathcal{K}^2\mathcal{H}^1=2(s_0\bar{s}_0)^{-1}(\hat{F}^-\cdot\hat{F}^--D^2)(\hat{F}^+\cdot\hat{F}^+-D^2)$. The operation $[~]_D$ defined in \eqref{eq:D_operation} immediately leads to $[\mathcal{R}]_D=\frac{e}{2}(\mathcal{R})_D=e(s_0\bar{s}_0)^{-1}(\hat{F}^-\cdot\hat{F}^--D^2)(\hat{F}^+\cdot\hat{F}^+-D^2)$. The FI Lagrangian \eqref{eq:bosonic_FI_lagrangian} is therefore given by:
\begin{equation}
e^{-1}\mathcal{L}_{FI}^{(\alpha_i)}=-\xi_is_0\bar{s}_0 e^{(\alpha_i+\frac{1}{3})K}(W\bar{W})^{\alpha_i+\frac{2}{3}}D.
\end{equation}
Since we are interested in matter coupled $\mathcal{N}=1$ supergravity, we use the Einstein frame where the conformal symmetry is gauge fixed through $s_0=\bar{s}_0=e^{\frac{K}{6}}$. This leads to a set of K\"ahler invariant terms parametrised by some constants $\{\alpha_i,\xi_i\}$ according to:
\begin{equation}\label{eq:FI_lagrangian_4}
e^{-1}\mathcal{L}_{FI}^{(\alpha_i)}=-\xi_ie^{\left(\alpha_i+\frac{2}{3}\right)K}(W\bar{W})^{\alpha_i+\frac{2}{3}}D=-\xi_ie^{\left(\alpha_i+\frac{2}{3}\right)\mathcal{G}}D\,,
\end{equation}
where $\mathcal{G}\equiv K+\ln |W|^2$.
Therefore, after gauge fixing the conformal symmetry and integrating out the auxiliary fields, the pure bosonic sector arising from the Lagrangian \eqref{eq:full_lagrangian} is given by
\begin{equation}
e^{-1}\mathcal{L}^{(B)}=\frac{1}{2}R-\frac{1}{4g^2}F_{\mu\nu}F^{\mu\nu}-G_{I\bar J}\partial X^I\cdot\partial \bar{X}^{\bar J}-\mathcal{V}
\end{equation}
with the scalar potential
\begin{equation}\label{eq:general_scalar_potential}
\mathcal{V}=e^{\mathcal{G}}\left(\partial_{I}\mathcal{G}~G^{I\bar J}~\partial_{\bar J}\mathcal{G}-3\right)+\mathcal{V}_{FI}.
\end{equation}
The new FI contribution to the scalar potential, $\mathcal{V}_{FI}$, arising from Eq.~\eqref{eq:FI_lagrangian_4} reads
\begin{equation}\label{eq:sum_FI_terms}
\mathcal{V}_{FI}=\frac{g^2}{2}\left(\underset{i}{\sum}\xi_ie^{\left(\alpha_i+\frac{2}{3}\right)\mathcal{G}}\right)^2,
\end{equation}
which is obviously K\"ahler invariant while field dependent at the same time. 

The above construction therefore provides a way to obtain an arbitrary set of (K\"ahler invariant) FI terms from a single $U(1)$ gauge field, in the presence of a superpotential $W$ with non-vanishing expectation value. Each term of the sum is parametrised by two real constants $\xi_i$ and $\alpha_i$. A constant FI term is obviously recovered by choosing one $\alpha_{i_0}=-{2}/{3}$. For this value, the bosonic part of the Lagrangian \eqref{eq:FI_lagrangian_3} is equal to the one of the Lagrangian \eqref{eq:FI_lagrangian_2}. Being independent of $W$, it is in particular valid even for vanishing superpotential, like the new FI term \eqref{eq:FI_lagrangian_2}. However, it is not clear that the fermionic parts of the Lagrangians \eqref{eq:FI_lagrangian_3} and \eqref{eq:FI_lagrangian_2} are equal, as well, for $\alpha_{i_0}=-{2}/{3}$.

A general sum appearing in the Lagrangian \eqref{eq:full_lagrangian}, involving terms of the form \eqref{eq:FI_lagrangian_4}, using that $e^{\mathcal{G}/2}=m_{3/2}[X]$, amounts to adding a general function of the gravitino mass $m_{3/2}[X]$ considered as a functional of the scalar fields $\{X\}$:
\begin{equation}
e^{-1}\mathcal{L}_{FI}^{(B)}=-f(m_{3/2}[X])D \quad\longrightarrow\quad \mathcal{V}_{FI}=\frac{g^2}{2} \mid f(m_{3/2}[X])\mid^2
\end{equation}
This construction allows us to refine the scalar potential by adding new field dependent and K\"ahler invariant terms. In the following, we will restrict ourselves as an illustration to the study of the simple case of one term of the type \eqref{eq:FI_lagrangian_4} up to an additive constant, corresponding to the choice $i=1,2$ with $\alpha_1$ an arbitrary parameter and $\alpha_2=-2/3$. Considering the K\"ahler potential of no-scale type and a constant superpotential, we will show that this choice is sufficient to produce inflationary models compatible with the slow-roll conditions and consistent with the CMB observations, with the inflaton rolling towards a de Sitter vacuum with tuneable energy and spontaneously broken supersymmetry.

\section{No-scale models and cosmological applications}
\label{sect:inflation}
In this section, we study the cosmological consequences of the previous modified FI-term construction in the case of simple no-scale models. Considering one chiral superfield $X$ associated to the inflaton, we successively choose the K\"ahler potentials
\begin{equation}
K(X,\bar X)=-\ln(X+\bar X)~~~\text{and}~~~K(X,\bar X)=-3\ln(X+\bar X),
\end{equation}
together with a constant superpotential $W=W_0$ and an exponential one $W(X)=e^{\beta X}$, respectively. In the context of string theory, these forms of K\"ahler potentials arise in all toroidal/orbifold compactifications as well as in the large volume limit of Calabi-Yau compactifications, both in heterotic string and in type II orientifolds. In this context, the first K\"ahler potential could describe for instance the kinetic term of the dilaton, associated to the string coupling, while the second may describe the internal volume of the $3$-complex dimensional compact space. We will therefore refer to the "dilaton case" and "compact volume case" to describe these two models. From now on, we also restrict the sum \eqref{eq:sum_FI_terms} to only two terms parametrised by three constants $\xi_1$, $\alpha_1\equiv\alpha$ and $\xi_2$, while $\alpha_2=-\frac{2}{3}$.

\subsection{Dilaton case}
\label{sect:dilaton_sector}
We first consider the K\"ahler potential $K=-\ln(X+\bar{X})$. In terms of the gravitino mass $m_{3/2}^2=e^{\mathcal{G}}$, this yields the scalar potential:
\begin{equation}\label{eq:scalar_potential_ungauged_shift_sym}
\mathcal{V}=-2m_{3/2}^2+\frac{1}{2}\left(\xi_1(m_{3/2}^2)^{\alpha+2/3}+\xi_2\right)^2\,,
\end{equation}
where we have redefined the parameters $\xi_i$ to absorb the gauge coupling constant $g$.
As we will show below, there is a region of the parameter space $\xi_1$, $\xi_2$ and $\alpha$, such that the above potential has an inflational plateau allowing slow-roll inflation compatible with the cosmological observations, and a minimum, where supersymmetry is spontaneously broken, with a tuneable vacuum energy by a fine tuning of the parameters (for instance to obtain a vanishing cosmological constant in the vacuum). 

In order to compute the slow-roll parameters, one needs to work with the canonically normalised field $\chi$, defined by its kinetic term through
\begin{equation}\label{eq:def_canonical_field}
\frac{\partial_{\mu}X\partial^{\mu}\bar{X}}{(X+\bar X)^2}=\frac{1}{2}\partial_{\mu}\chi\partial^{\mu}\chi+...
\end{equation}
where the dots denote terms containing the imaginary part of $X$, which has no influence on the discussion of this section. We will come back to it in Section \ref{sect:gauging_shift_symmetry}, where the shift symmetry associated to this imaginary part will be gauged by the $U(1)$. Focusing on the real part for now, we deduce from \eqref{eq:def_canonical_field}
\begin{equation}\label{dcX}
{\rm Re} X=e^{\sqrt 2 \chi},    
\end{equation}
and thus 
\begin{equation}\label{eq:Lagrangian_gravitino_mass}
m_{3/2}^2=e^{\mathcal G}=\frac{|W_0|^2}{2}e^{-\sqrt 2 \chi}.
\end{equation}

In the following, the `dilaton' $\chi$ will be identified with the inflaton, dynamically driving inflation starting from a large value, slightly rolling down along the potential, attaining the horizon exit denoted by $\chi_*$ and ending at a value $\chi_{\text{end}}$ when slow-roll stops. The field then continues to fall down towards the minimum, when reheating takes place. From now on, quantities observed at the horizon exit are specified with a star *, and the approximation of large inflaton field $\chi>>1$ is assumed in this region\footnote{$\chi>>1$ corresponds to                                                                                                                                                                                                                                                                                                                                                                                                                                                                                                                                                                                                                                                                                                                                                                                                                                                                                                                                                                                                                                                                                                                                                                                                                                                                                                                                                                                                                                                                                                                                                                                                                                                                                                                                                                                                                                                                                                                                                                                                                                                                                                                                                                                                                                                                                                                                                                                                                                                                                                                                                                                                                                                                                                                                                                                                                                                                                                                                                                                                                                                                                                                                                                                                                                                                                                                                                                                                                                                                                                                                                                                                                                                                                                                                                                                                                                                                                                                                                                                                                                                                                                                                                                                                                                                                                                                                                                                                                                                                                                                                                                                                                                                                                                                                                                                                                                                                                                                                                                                                                                                                                                                                                                                                                                                                                                                                                                                                                                                                                                                                                                                                                                                                                                                                                                                                                                                                                                                                                                                                                                                                                                                                                                                                                                                                                                                                                                                                                                                                                                                                                                                                                                                                                                                                                                                                                                                                                                                                                                                                                                                                                                                                                                                                                                                                                                                                                                                                                                                                                                                                                                                                                                                                                                                                                                                                                                                                                                                                                                                                                                                                                                                                                                                                                                                                                                                                                                                                                                                                                                                                                                                                                                                                                                                                                                                                                                                                                                                                                                                                                                                                                                                                                                                                                                                                                                                                                                                                                                                                                                                                                                                                                                                                                                                                                                                                                                                                                                                                                                                                                                                                                                                                                                                                                                                                                                                                                                                                                                                                                                                                                                                                                                                                                                                                                                                                                                                                                                                                                                                                                                                                                                                                                                                                                                                                                                                                                                                                                                                                                                                                                                                                                                                                                                                                                                                                                                                                                                                                                                                                                                                                                                                                                                                                                                                                                                                                                                                                                                                                                                                                                                                                                                                                                                                                                                                                                                                                                                                                                                                                                                                                                                                                                                                                                                                                                                                                                                                                                                                                                                                                                                                                                                                                                                                                                                                                                                                                                                                                                                                                                                                                                                                                                                                                                                                                                                                                                                                                                                                                                                                                                                                                                                                                                                                                                                                     weak coupling, which is necessary for the validity of an effective supergravity theory. However, the large field approximation is not really needed; instead, the required condition is that $m_{3/2}\to 0$.}. 
The gravitino mass~\eqref{eq:Lagrangian_gravitino_mass} vanishes exponentially and the potential~\eqref{eq:scalar_potential_ungauged_shift_sym} for $\alpha\ge -2/3$ is therefore dominated by a constant, as required by slow-roll inflation. In the following, we will thus restrict to the region $\alpha> -2/3$, with $\mathcal{V}_*\simeq\frac{\xi_2^2}{2}$.
In terms of the canonical field $\chi$, the slow-roll parameters are given as usual by 
\begin{equation}
\epsilon\equiv\frac{1}{2}\left(\frac{\partial V/\partial \chi}{V}\right)^2\quad;\quad \eta\equiv\frac{\partial^2V/\partial\chi^2}{V}\,.
\end{equation}
At large field $\chi$, by further assuming $\alpha>1/3$, they can be expanded into 
\begin{equation}\label{eq:slow_roll_param_dilaton_case}
\epsilon\underset{\chi>>1}{\approx}\left(\frac{4m_{3/2}^2}{\xi_2^2}\right)^2\underset{\chi>>1}{\approx}\frac{1}{4}\eta^2\quad;\quad \eta\underset{\chi>>1}{\approx}-\frac{8m_{3/2}^2}{\xi_2^2}\,.
\end{equation}
Actually, the large field condition is not really necessary. The required approximation is that the gravitino mass~\eqref{eq:Lagrangian_gravitino_mass} should be small during inflation, so  that the potential ~\eqref{eq:scalar_potential_ungauged_shift_sym} is approximately constant. This is an important point, implying that the models we study are consistent with small field inflation, avoiding trans-planckian initial conditions for the normalised inflaton field.

A central quantity to be taken into account in inflation is the number $N$ of e-folds between the horizon exit and the end of inflation, a period observable through the CMB. This quantity, which must be set within the range $[40,60]$ to satisfy CMB observations, is given by:
\begin{equation}\label{eq:e_folds_number}
N=\int_{\chi_*}^{\chi_{\text{end}}}\frac{d\chi}{\sqrt{2\epsilon(\chi)}}\in[40,60].
\end{equation}
Two other observable quantities at the horizon exit are the amplitude of primordial density fluctuations $A_S$ and the spectral index, or tilt $n_s$, respectively given by
\begin{eqnarray}
\label{eq:amplitude}
A_S&=&\frac{\mathcal{V}_*}{24\pi^2\epsilon_*}=2.2\times 10^{-9},\\
\label{eq:tilt}
n_S&=&1+2\eta_*-6\epsilon_*=0.96,
\end{eqnarray}
where the numerical equalities also follow from the CMB data. 

To be consistent with observations, the inflaton potential during inflation should respect the three conditions \eqref{eq:e_folds_number}, \eqref{eq:amplitude} and \eqref{eq:tilt}, which we now use in order to constrain the three parameters $\xi_1$, $\xi_2$ and $\alpha$. In the large field limit at the horizon exit, the tilt \eqref{eq:tilt} gives $n_S=1+2\eta_*-6\epsilon_*\simeq 1+2\eta_*-\frac{3}{2}\eta_*^2\approx 1+2\eta_*=1-{16m_{3/2}^{*2}}/{\xi_2^2}\simeq 0.96$, from which we deduce that $\xi_2^2\simeq 400 m_{3/2}^{*2}$. Moreover, the amplitude \eqref{eq:amplitude} leads to $\frac{1}{768\pi^2}\frac{\xi_2^6}{m_{3/2}^{*4}}=2.2\times 10^{-9}$. Therefore, in the large field limit, the $\xi_1$ and $\alpha$ dependence drops, and one can immediately find from these two relations the numerical values for the parameter $\xi_2$ and the gravitino mass at the horizon exit $m_{3/2}^{*2}$, namely:
\begin{eqnarray}
\label{eq:xi_2}\xi_2^2&=&1.04\times10^{-10},\\
\label{eq:gravitino_mass_horizon}m_{3/2}^{*2}&=&2.6\times10^{-13}.
\end{eqnarray}
From the value of $\epsilon$ at the horizon, given by the first equation of \eqref{eq:slow_roll_param_dilaton_case}, and the two relations \eqref{eq:xi_2} and \eqref{eq:gravitino_mass_horizon}, we find the predicted value for the tensor-to-scalar ratio of primordial perturbations to be:
\begin{equation}r=16\epsilon_*\simeq\left(\frac{16m_{3/2}^{*2}}{\xi_2^2}\right)^2=1.6\times 10^{-3},
\end{equation}
which is fixed and independent of any parameters of the model, as long as $\alpha>1/3$ is considered.

On the other hand, the condition to have a (almost) vanishing potential at its minimum\footnote{This is obviously an approximation since the cosmological constant is extremely small but nonzero. But the point to keep in mind is that in this model, the cosmological constant can be tuned to \emph{any} small positive value.}, for a value of the gravitino mass denoted $\tilde m_{3/2}^2$ in what follows, can then be used in order to determine the parameter $\xi_1$ in terms of $\alpha$. This is obtained by numerically solving $\mathcal{V}(\tilde m_{3/2}^2)=0$, with the result denoted $\xi_1(\alpha)$ in the following. 
In order to constrain the last remaining parameter $\alpha$ via the number of e-folds equation \eqref{eq:e_folds_number}, we first need to determine the value of the inflaton field at the end of inflation, depending on $\alpha$. Inflation stops when $\chi$ reaches a value $\chi_{\text{end}}$ such that $\epsilon(\chi_{\text{end}})=1$ or $|\eta(\chi_{\text{end}})|=1$. In this model, the condition first fulfilled turns out to be $\eta(\chi_{\text{end}})=-1$, which leads to the equation:
\begin{equation}\label{eq:chi_end_1} 
6m_{3/2}^{2~\text{end}}-\xi_1^2(\alpha)(m_{3/2}^{2~\text{end}})^{2\alpha+4/3}\left(\frac{1}{2}+4\left(\alpha+\frac{2}{3}\right)^2\right)-\xi_1(\alpha)\xi_2(m_{3/2}^{2~\text{end}})^{\alpha+2/3}\left(1+2\left(\alpha+\frac{2}{3}\right)^2\right)-\frac{\xi_2^2}{2}=0.
\end{equation}
This equation is solved numerically to get $m_{3/2}^{2~\text{end}}$ in terms of $\alpha$. The number of e-folds is then used in order to determine the parameter $\alpha$. Indeed, equation \eqref{eq:e_folds_number} becomes:
\begin{equation}
N(\alpha)=-\frac{1}{4}\int_{m_{3/2}^{*2}}^{m_{3/2}^{2~\text{end}}(\alpha)}\frac{-4m_{3/2}^2+\left[\xi_1(\alpha)(m_{3/2}^2)^{\alpha+2/3}+\xi_2\right]^2}{2m_{3/2}^2-(\alpha+\frac{2}{3})\xi_1^2(\alpha)(m_{3/2}^2)^{2\alpha+4/3}-(\alpha+\frac{2}{3})\xi_1(\alpha)\xi_2(m_{3/2}^2)^{\alpha+2/3}}\,\frac{dm_{3/2}^2}{m_{3/2}^2}\,.
\end{equation}
Using $m_{3/2}^{2~\text{end}}$ given by the largest solution of Eq.~\eqref{eq:chi_end_1}, the value  for $\xi_2^2$ \eqref{eq:xi_2}, and the expression for $\xi_1(\alpha)$ given from the solution of $\mathcal{V}(\tilde m_{3/2}^2)=0$, the above integral can be numerically evaluated in terms of $\alpha$. It turns out that any $\alpha$ larger or approximately equal to $1$ leads to an acceptable e-fold number $N\in[40,60]$. Thus, the only fine tuning of the model, besides fixing the overall scale of the potential by its asymptotic value determined by $\xi_2$, is related to the vacuum energy. The gravitino mass at the minimum of the potential $\tilde{m}_{3/2}$ can be between $10^{10}$TeV and the Planck scale by choosing $\alpha$ between $1$ and $10.5$ respectively.

\begin{figure}[!h]
    \begin{minipage}[c]{.46\linewidth}
        \centering
        \includegraphics[height=50mm]{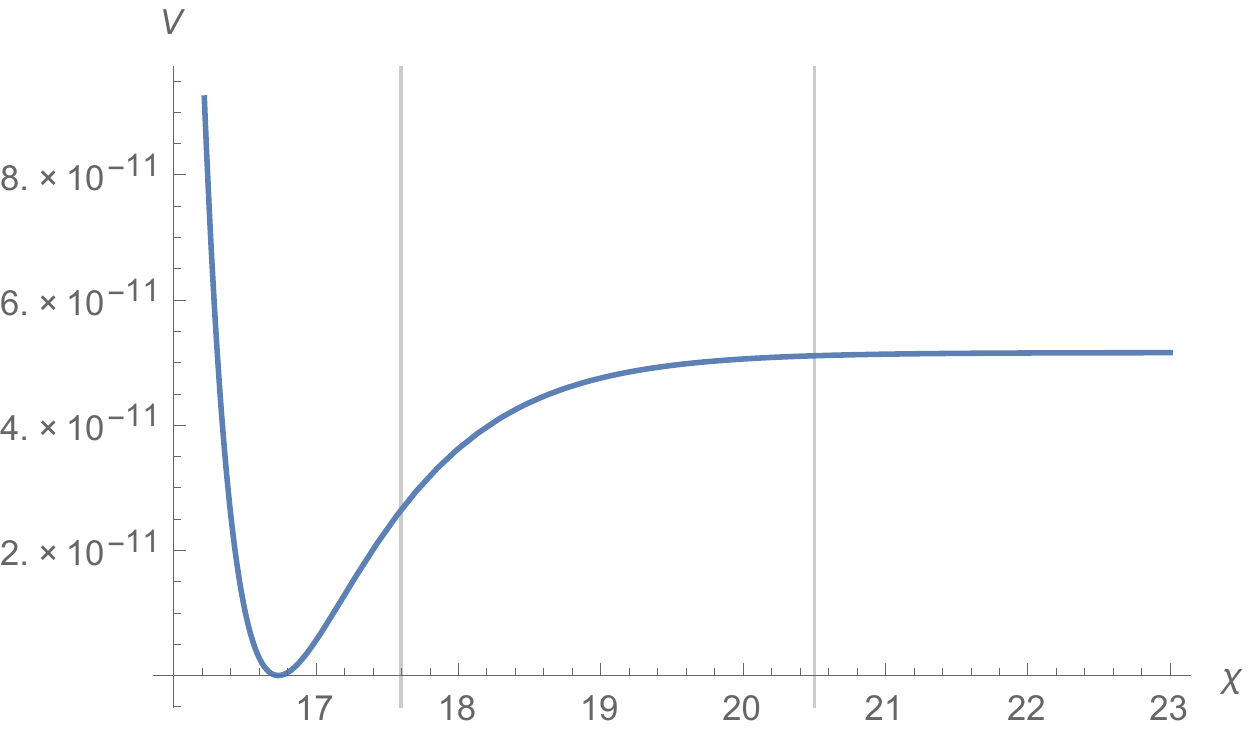}
        \caption{Scalar potential as a function of the canonically normalised field $\chi$, for $\alpha=1$}
	\label{fig:Scalar potential}
    \end{minipage}
    \hfill
    \begin{minipage}[c]{.46\linewidth}
        \centering
        \includegraphics[height=50mm]{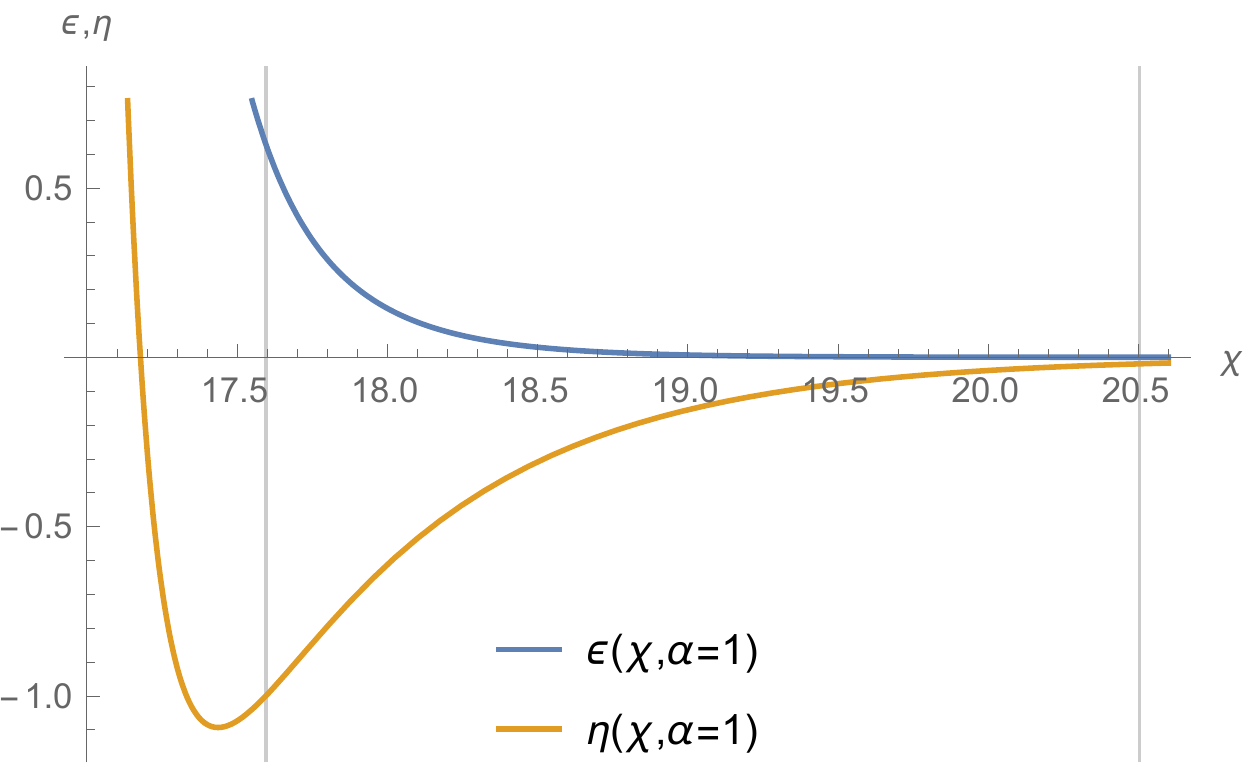}
        \caption{Slow-roll parameters $\epsilon$ and $\eta$ as a function of $\chi$, for $\alpha=1$}
	\label{fig:Slow-roll parameters}
    \end{minipage}
\end{figure}

As an illustration, we now choose $\alpha\simeq 1$, which gives $\xi_1(\alpha=1)\simeq 10^{11}$.\footnote{Despite the large value of $\xi_1$, one can check that the approximation $\mathcal{V}_{*}={\xi_2^2}/{2}$ at the horizon, assumed in the computation of the tilt and of the amplitude, is valid. Indeed, with the numerical values $\alpha\sim 1$, $\xi_1\sim 10^{11}$, $\xi_2\sim 10^{-5}$ and $m_{3/2}^{*2}=2.64\times 10^{-13}$, we get $\frac{\xi_1(m_{3/2}^{*2})^{\alpha+2/3}}{\xi_2}\sim 6\times 10^{-5}$.} With these values, the scalar potential and the slow-roll parameters are plotted in terms of $\chi$ in Figures \ref{fig:Scalar potential} and \ref{fig:Slow-roll parameters} respectively, where we have also set $W_0=\sqrt{2}$. The vertical grey lines indicate the horizon exit and the end of inflation (from the right to the left). The corresponding numerical values for the gravitino mass are $m_{3/2}^{*2}=2.64\times 10^{-13}$ and $m_{3/2}^{2\text{end}}=1.56\times 10^{-11}$ in Planck units. The minimum is reached at $\tilde m_{3/2}^2=5.29\times 10^{-11}$. The associated values for the inflaton field are $\chi_*=20.48+\frac{1}{\sqrt{2}}\ln{\frac{|W_0|^2}{2}}$, $\chi_{\text{end}}=17.59+\frac{1}{\sqrt{2}}\ln{\frac{|W_0|^2}{2}}$ and $\tilde{\chi}=16.73+\frac{1}{\sqrt{2}}\ln{\frac{|W_0|^2}{2}}$. Notice that the values of the inflaton can be made less than one for an appropriate choice of $W_0$, as we already mentioned in the begining of the section. Finally, note that because of the space-time curvature during inflation, the value of $m_{3/2}$ entering in the Lagrangian is not the physical gravitino mass, which should be computed taking into account the curvature contribution in an approximate de Sitter spacetime~\cite{mass_de_sitter_space_1,mass_de_sitter_space_2}. 

The spectrum at the minimum contains the imaginary part of $X$ and the $U(1)$ gauge boson, which remain massless in this model, as it can be seen from the expression of the scalar potential, as well as the massive gravitino and inflaton whose masses are given by:
\begin{align}
\tilde m_{3/2}^2 &= 5.29\times 10^{-11}, &
\tilde m_{\chi}^2 &= \left.\frac{\partial^2V}{\partial\chi^2}\right|_{\chi=\chi_{\text{min}}}=2.46\times10^{-10}.
\end{align}
There is also a massive spin-1/2 fermion corresponding to a linear combination of the $U(1)$ gaugino and the fermionic component of the inflaton superfield, orthogonal to the Goldstino direction. Indeed at the minimum, supersymmetry is spontaneously broken by a non-vanishing expectation value of both a D and F-term. The Goldstino $P_L\nu$ is thus a linear combination of the gaugino $\lambda$ and of the chiral fermion $\Omega$: $P_L\nu=\frac{1}{\sqrt 2}\Omega^Xg_{X\bar X}\bar{F}^{\bar X}-\frac{i}{2}DP_L\lambda$, with $\bar{F}^{\bar X}\equiv-e^{K/2}g^{X\bar X}\nabla_X W$, evaluated at the minimum. In order to compute the direction of supersymmetry breaking, we consider:
\begin{eqnarray}
\label{eq:F_term}
\| F\|&\equiv&\sqrt{F^Xg_{X\bar X}\bar{F}^{\bar X}}=\sqrt{e^{\mathcal{G}}\partial_X\mathcal{G}G^{X\bar X}\partial_{\bar{X}}\mathcal{G}}=m_{3/2},\\
\label{eq:D_term}
\| D\|&\equiv&\xi_1e^{5/3\mathcal{G}}+\xi_2=\xi_1(m_{3/2}^2)^{5/3}+\xi_2.
\end{eqnarray}
At the minimum, we have:
\begin{equation}
\left\| \frac{D}{F}\right\|_{\tilde{m}_{3/2}}=\left[\xi_1(\tilde m_{3/2}^2)^{7/6}+\xi_2(\tilde m_{3/2}^2)^{-1/2}\right]\simeq 1.5\,,
\end{equation}
where we have used the values $\xi_1=10^{11}$, $\xi_2=10^{-5}$ and $\tilde m_{3/2}^2=5.29\times10^{-11}$ obtained previously. At the minimum, the Goldstino is thus an approximately equal mixing of the chiral fermion $\Omega$ and the gaugino $\lambda$. \\
The computation of the fermion masses is detailed in Appendix \ref{sect:append_fermion_masses}. The mass squared $m_f^2$ of the physical fermion which remains after elimination of the Goldstino is given in Eq.~\eqref{eq:physical_fermion_mass}. For $p=1$, its numerical value at the minimum where $\tilde{m}_{3/2}^2=5.29\times 10^{-11}$ is (in Planck units):
\begin{equation}
m_f^2=5.9\times 10^{-12}.
\end{equation}

\subsection{Compact volume case}
\label{sect:volume_sector}
In this subsection, we consider the no-scale model with K\"ahler potential $K(X,\bar X)=-3\ln(X+\bar X)$. If one takes a constant superpotential as in the previous subsection, the F-term of the scalar potential will vanish, and the new Fayet-Iliopoulos term will be ill-defined at the minimum, where $D$ now vanishes\footnote{Of course, the vanishing of the F-part of the scalar potential is a tree-level result and can be circumvented by considering quantum corrections in the K\"ahler potential.}. Instead, we consider a superpotential of the form $W(X)=e^{\beta X}$, with $\beta$ a real constant. Note that the the imaginary shift of $X$ becomes now a (global) R-symmetry~\cite{Antoniadis:2017gjr}. The full scalar potential is then given by:
\begin{equation}\label{cvc_full_potential}
\mathcal{V}=m_{3/2}^2\left[-3+\frac{1}{3}\left(\beta(X+\bar X)-3\right)^2\right]+\frac{1}{2}\left(\xi_1(m_{3/2}^2)^{\alpha+2/3}+\xi_2\right)^2.
\end{equation}
Choosing $\beta<<\left.(X+\bar X)^{-1}\right|_{*}$, the first term of \eqref{cvc_full_potential} can be neglected at the horizon exit as well as during the inflationary period. However, outside of the inflationary plateau, the D-term starts decreasing significantly and the F-term cannot be neglected anymore. Supersymmetry at the minimum of the potential is then spontaneously broken by non-vanishing expectation values of both D and F-terms, and a tuning of the parameters would be required in order to get a vanishing potential at its minimum, as in the previous case studied above. We will not study this region in the following, focusing on the inflationary period where the F contribution to $\mathcal{V}$ can be neglected and the scalar potential is only given by its D-term:
\begin{equation}\label{cvc_potential}
\left.\mathcal{V}\right|_{\text{infla.}}=\frac{1}{2}\left(\xi_1(m_{3/2}^2)^{\alpha+2/3}+\xi_2\right)^2\,.
\end{equation}

Now the normalised field $\chi$ and the gravitino mass are given by: 
\begin{equation}
{\rm Re}X=e^{\sqrt{\frac{2}{3}}\chi} \quad;\quad m_{3/2}^2=\frac{|W|^2}{(X+{\bar X})^3}=\frac{|W|^2}{8}e^{-\sqrt{6}\chi}\,.
\end{equation}
Like in the previous subsection, the potential at the horizon exit, where $\chi>>1$ is assumed\footnote{$\chi>>1$ now corresponds to a large volume of the compact space, which is compatible with the effective theory where higher derivatives are neglected.}, is given by $\mathcal{V}_*={\xi_2^2}/{2}$. The slow-roll parameters expanded in this limit read: 
\begin{equation}\label{eq:slow_roll_param_compact_case}
\epsilon\underset{\chi>>1}{\approx}\frac{12\xi_1^2(\alpha+\frac{2}{3})^2(m_{3/2}^2)^{2\alpha+4/3}}{\xi_2^2}\quad;\quad\eta\underset{\chi>>1}{\approx}\frac{12\xi_1(\alpha+\frac{2}{3})^2(m_{3/2}^2)^{\alpha+2/3}}{\xi_2}\,,
\end{equation}
and thus $\eta^2\underset{\chi>>1}{\approx}12(\alpha+\frac{2}{3})^2\epsilon$. With these two quantities, the tilt and amplitude analysis yields:
\begin{align}\label{eq:gravitino_mass_horizon_compact_sector}
\xi_2(\alpha) = \frac{6\times10^{-6}}{\alpha+2/3}\quad,\quad 
(m_{3/2}^{*2})^{\alpha+2/3}(\xi_1,\alpha) &= -\frac{1}{\xi_1}\,\frac{1.02\times10^{-8}}{(\alpha+2/3)^3}\,.
\end{align}

The gravitino mass at the end of inflation, $m_{3/2}^{2~\text{end}}$, is still given by the condition $\eta(m_{3/2}^{2~\text{end}})=\pm 1$, which is now solution of the equation:   
\begin{equation}
\xi_1^2\,(m_{3/2}^{2~\text{end}})^{2\alpha+4/3}\left(\frac{1}{2}\mp 12\left(\alpha+\frac{2}{3}\right)^2\right)+\xi_1\,\xi_2(\alpha)\,(m_{3/2}^{2~\text{end}})^{\alpha+2/3}\left(1\mp 6\left(\alpha+\frac{2}{3}\right)^2\right)+\frac{\xi_2(\alpha)^2}{2}=0\,.
\end{equation}
This can be solved analytically at fixed $\alpha$, yielding:
\begin{equation}\label{eq:gravitino_mass_end}
(m_{3/2}^{2~\text{end}})^{\alpha+2/3}_{\pm}(\xi_1,\alpha)=-\frac{\xi_2(\alpha)}{\xi_1}\times\frac{1\mp 6(\alpha+2/3)^2-2\sqrt{3}(\alpha+2/3)\sqrt{3(\alpha+2/3)^2\pm 1}}{1\mp 24(\alpha+2/3)^2}\,.
\end{equation}
On the other hand, the number of e-folds is given by:
\begin{equation}
N_{\pm}(\alpha)=\frac{-1}{12(\alpha+\frac{2}{3})}\int_{m_{3/2}^{*2}(\xi_1,\alpha)}^{{m_{3/2}^{2~\text{end}}}_{\pm}(\xi_1,\alpha)}\frac{\left[\xi_1(m_{3/2}^2)^{\alpha+2/3}+\xi_2(\alpha)\right]^2}{\xi_1^2(m_{3/2}^2)^{2\alpha+4/3}+\xi_1\xi_2(\alpha)(m_{3/2}^2)^{\alpha+2/3}}\, \frac{dm_{3/2}^2}{m_{3/2}^2}\,,
\end{equation}
which is independent of $\xi_1$, as can be seen from the change of variable $m_{3/2}^2\rightarrow m_{3/2}^2\xi_1^{\frac{1}{\alpha+2/3}}$ and by using the second equation of \eqref{eq:gravitino_mass_horizon_compact_sector} and Eq.~\eqref{eq:gravitino_mass_end}. 

\begin{figure}[!h]
    \begin{minipage}[c]{.46\linewidth}
        \centering
        \includegraphics[height=50mm]{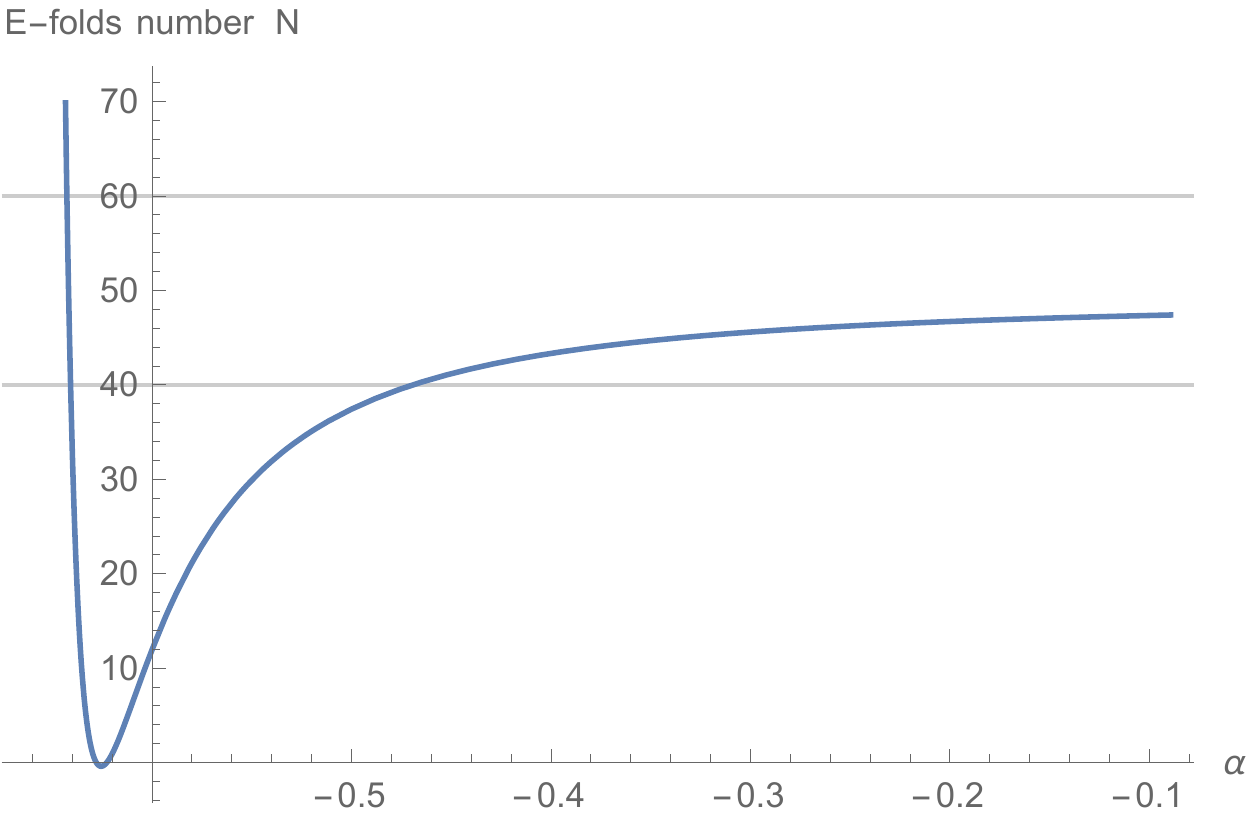}
        \caption{Number of e-folds $N_{+}$ as a function of $\alpha$, for $-\frac{2}{3}<\alpha<\frac{\sqrt{3}-2}{3}$}
	\label{fig:e_folds_number_1}
    \end{minipage}
    \hfill
    \begin{minipage}[c]{.46\linewidth}
        \centering
        \includegraphics[height=50mm]{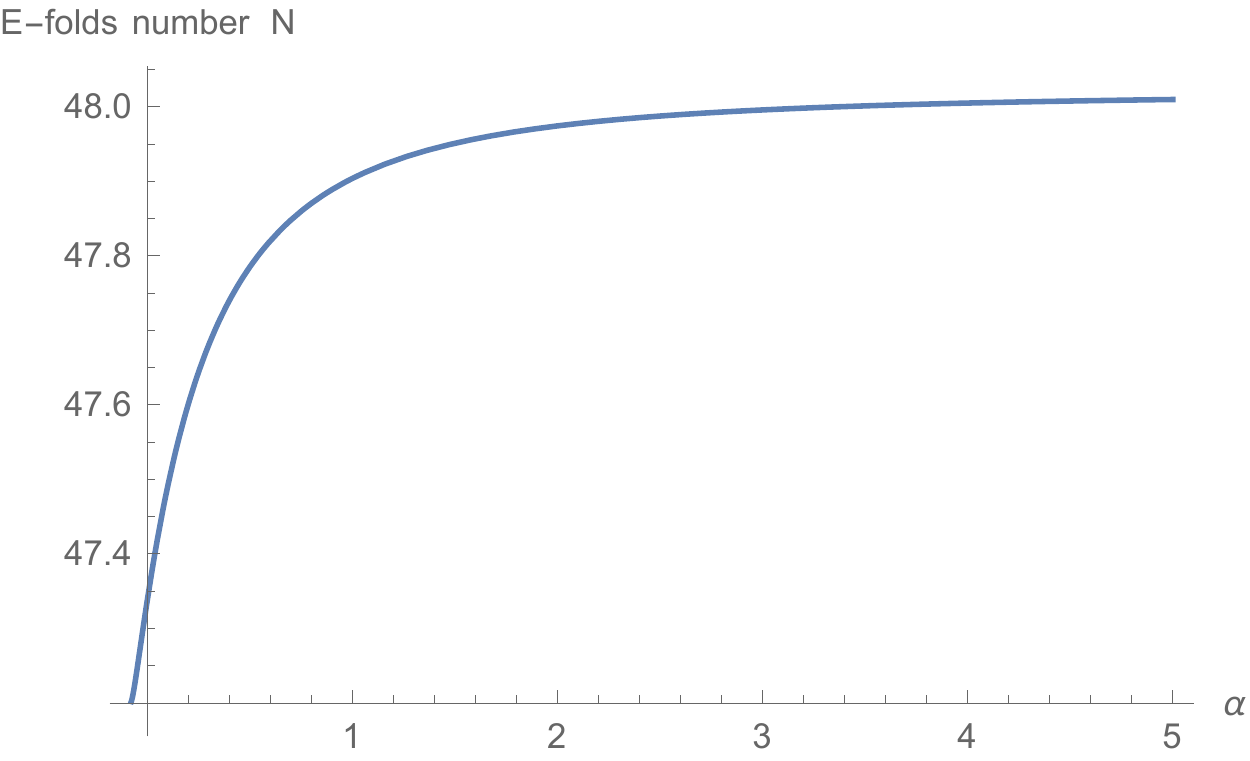}
        \caption{Number of e-folds $N_{-}$ as a function of $\alpha$, for $\alpha\geq \frac{\sqrt{3}-2}{3}$}
	\label{fig:e_folds_number_2}
    \end{minipage}
\end{figure}

Two regions for $\alpha$ have to be considered: (i) $-\frac{2}{3}<\alpha<\frac{\sqrt{3}-2}{3}$, where $\eta(m_{3/2}^{2~\text{end}})=1$ is first fulfilled, and where the gravitino mass at the end of inflation and the number of e-folds are respectively given by $(m_{3/2}^{2~\text{end}})_{+}$ and $N_{+}$; (ii) $\alpha\geq\frac{\sqrt{3}-2}{3}$, where $\eta(m_{3/2}^{2~\text{end}})=-1$ is first fulfilled, and where the gravitino mass at the end of inflation and the number of e-folds are respectively given by $(m_{3/2}^{2~\text{end}})_{-}$ and $N_{-}$. Both e-fold numbers are plotted in terms of $\alpha$ in Figures \ref{fig:e_folds_number_1} and \ref{fig:e_folds_number_2}. Any $\alpha\gtrsim-0.46$ leads to an acceptable $N\in[40,60]$, while the parameter $\xi_1$ remains undetermined.
From the value of $\epsilon$ at the horizon given by the first equation of \eqref{eq:slow_roll_param_compact_case} and the two relations \eqref{eq:gravitino_mass_horizon_compact_sector}, one sees that the predicted value for the tensor-to-scalar ratio $r$ of primordial perturbations remains independent of $\xi_1$, and depends only on $\alpha$:
\begin{equation}
r(\alpha)=16\,\epsilon_*(\alpha)=16\frac{12(\alpha+2/3)^2\xi_1^2(m_{3/2}^{*2})^{2\alpha+4/3}(\xi_1,\alpha)}{\xi_2^2(\alpha)}\simeq \frac{5.4\times 10^{-4}}{(\alpha+2/3)^2}\,.
\end{equation}
Thus, $\alpha$ can be chosen such that the tensor-to-scalar ratio is large and close to the experimental bound, for instance $r(\alpha=-0.45)\simeq 10^{-2}$ with $N_{+}(\alpha=-0.45)\simeq 41$. 

\section{Gauging the axion shift symmetry}
\label{sect:gauging_shift_symmetry}
In the two previous models, the spectrum contained two massless particles: the imaginary part of the complex inflaton field $X$, and the $U(1)$ gauge boson $A_{\mu}$, which is an unwanted phenomenological property. This can be avoided by gauging the imaginary shift symmetry by the $U(1)$. Under a gauge transformation $A_{\mu}\rightarrow A_{\mu}-2\partial_{\mu}\lambda$, one then has for the complex scalar $X\rightarrow X+ic\lambda$, with $\lambda$ the gauge parameter and $c$ a constant related to the charge $e^c$ of the field $e^X$. In terms of superfields, this transformation reads $X\rightarrow X+c\Lambda$, with $\Lambda$ a chiral superfield gauge parameter. The gauge transformation of the vector superfield $V$ is $V\rightarrow V-\Lambda-\bar{\Lambda}$. In order to keep a gauge invariant K\"ahler potential with shift symmetry, $K(X+\bar X)$ must be modified as:
\begin{equation}\label{eq:new_Kahler potential0}
K(X+\bar X)\rightarrow K(X+\bar X+cV)\,.
\end{equation}
Note that this modification does not change the pure bosonic part of the FI Lagrangian \eqref{eq:FI_lagrangian_3}. Indeed, when fermions are put to zero, the only non-vanishing components of the chiral multiplets $\bar{\lambda}P_L\lambda$ and $\bar{\lambda}P_R\lambda$ are their $\theta\theta$ and $\bar{\theta}\bar{\theta}$ components. Therefore, only the lowest components of the other superfields involved in \eqref{eq:FI_lagrangian_3} contribute to the bosonic sector, and the lowest component of $e^{(\alpha+\frac{1}{3})K}$ does not receive additional contributions from $cV$ in the Wess-Zumino gauge. 

In order to see how a massive gauge boson arises from this gauging, we work in global supersymmetry and compute the (bosonic) new terms appearing from this modification. Putting fermions to zero and expanding in components, we have
\begin{equation}
\left.X+\bar X+cV\right|_{\text{bos}}=2{\rm Re}X-\theta\sigma^{\mu}\bar{\theta}(cA_{\mu}+2\partial_{\mu}{\rm Im}X)+\frac{1}{2}\theta^2\bar{\theta}^2(cD-\partial^2{\rm Re}X)+\theta^2 F+\bar{\theta}^2\bar F,
\end{equation}
from which we deduce:
\begin{equation}
\left. K(X+\bar X+cV)\right|_{\theta^2\bar{\theta}^2}=\frac{K'}{2}(cD-\partial^2{\rm Re}X)-\frac{K''}{4}(cA_{\mu}+2\partial_{\mu}{\rm Im}X)^2+K''F\bar{F}.
\end{equation}
It follows that
\begin{equation}
\int d^4\theta K(X+\bar X+cV)=\int d^4\theta K(X+\bar X) 
-\frac{c^2}{4} K''A_{\mu}A^{\mu}
-cK''A_{\mu}\partial^{\mu}{\rm Im}X
+\frac{c}{2}K'D+\, {\rm fermions}\,.
\end{equation}
As a result, there is a mass term for the gauge boson $A_{\mu}$, as well as a new field dependent FI term $-\xi(X)D$, with $\xi(\chi)=-cK'/2$. It modifies the D-term of the scalar potential \eqref{eq:sum_FI_terms} according to 
$D=g^2\left[-cK'/2+\underset{i}{\sum}\xi_ie^{(\alpha_i+2/3)\mathcal{G}}\right]$, which leads to the following D-term contribution to the scalar potential:
\begin{equation}\label{eq:scalar_potential_gauged_shift_sym}
\mathcal{V}_{FI}=\frac{g^2}{2}\left(\underset{i}{\sum}\xi_ie^{\left(\alpha_i+\frac{2}{3}\right)\mathcal{G}}-\frac{c}{2}K'\right)^2.
\end{equation}

It is easy to show that the extra contribution proportional to $c$, due to the gauging of the shift symmetry, does not alter the inflationary predictions discussed in the previous section, when restricting the D-auxiliary field to only two non-vanishing terms, as in the previous section. Consider for example the compact volume case with a K\"ahler potential $K=-3\ln (X+{\bar X})$. The second term in \eqref{eq:scalar_potential_gauged_shift_sym} then becomes proportional to $m_{3/2}^{2/3}$ which may be identified as a particular case of the potential \eqref{cvc_potential} studied before for $\alpha=-1/3$ and $\xi_1=(3cg)/(2|W_0|^{2/3})$. $\xi_2$ can then be obtained from the first sum in \eqref{eq:scalar_potential_gauged_shift_sym} by choosing one non-vanishing term, say $\xi_2$ with $\alpha_2=-2/3$. The analysis is then reduced to the one of the last section, in the compact volume case with a fixed value of the parameter $\alpha=-1/3$, which is within the allowed region of the parameter space compatible with observational data, as seen in Fig.~\ref{fig:e_folds_number_1}. 

Let us finally consider another example inspired by the heterotic string with $X$ identified with the string dilaton, as in the first model considered in the previous section, where its axionic imaginary part is dual to the Neveu-Schwarz antisymmetric tensor in four dimensions. In this case, the constant $c$ is related to a $U(1)$ anomaly which is cancelled by a Green-Schwarz term. The gauging of the shift symmetry is a consequence of the anomaly cancellation and the axion is absorbed by the $U(1)$ becoming massive and no massless particle remains in the spectrum~\cite{GS}. The gauge coupling is not anymore constant but is fixed by $X$: $g^2=1/{\rm Re}X$, corresponding to a gauge kinetic function linear in $X$. The scalar potential can be easily obtained from Eq.~\eqref{eq:scalar_potential_gauged_shift_sym} using $K=-\ln (X+{\bar X})$ and the expressions \eqref{dcX} and \eqref{eq:Lagrangian_gravitino_mass}:
\begin{equation}
\mathcal{V}=-2m_{3/2}^2+\frac{e^{-\sqrt{2}\chi}}{2}\left(\underset{i}{\sum}\xi_ie^{\left(\alpha_i+\frac{2}{3}\right)\mathcal{G}}+\frac{c}{4}e^{-\sqrt{2}\chi}\right)^2.
\end{equation}
Again we restrict the D-term to only two non-vanishing contributions. In order to get an asymptotically constant potential at infinity, we choose $\alpha_2=-\frac{7}{6}$, while $\alpha_1=\frac{1}{3}$ is chosen to be able to absorb the constant $c$ in $\xi_1$. We obtain in this way a potential with the same form as in Eq.~\eqref{eq:scalar_potential_ungauged_shift_sym}:
\begin{equation}
\mathcal{V}=-2m_{3/2}^2+\frac{1}{2}\left(\xi_1'(m_{3/2}^2)^{3/2}+\xi_2'\right)^2,
\end{equation}
where we have defined $\xi_1'\equiv\frac{\xi_1\sqrt{2}}{W_0}+\frac{c}{\sqrt{2}W_0^3}$ and $\xi_2'\equiv\frac{\xi_2\sqrt{2}}{W_0}$.

The potential is thus the same as the one of Eq.~\eqref{eq:scalar_potential_ungauged_shift_sym}, with $\alpha=\frac{5}{6}$. This is an acceptable value since it leads to a number of e-folds $N(5/6)\simeq 51$. The 
numerical predictions obtained in Section \ref{sect:dilaton_sector} are not modified by the gauging of the shift symmetry. The main improvement is that now the imaginary part of the inflaton has been absorbed by the $U(1)$ gauge boson which acquires a mass. In order to compute this mass, one needs to rescale $A_{\mu}\rightarrow gA_{\mu}$ so that the gauge field kinetic term becomes canonical. After this rescaling, the gauge boson mass square reads: 
\begin{equation}
m_{A}^2(\chi)=\frac{e^{-3\sqrt 2\chi}}{8}c^2=\frac{g^6}{8}c^2.    
\end{equation}
With the values of $\chi$ at the horizon crossing and at the minimum found above, we get $m_{A}^2$ in terms of the parameters $c$ and $W_0$:
\begin{eqnarray}
m_A^{*2}&=&1.84\times 10^{-38}\frac{c^2}{W_0^6}\,,\\
m_A^{2\text{end}}&=&3.05\times 10^{-34}\frac{c^2}{W_0^6}\,,\\
\tilde m_A^2&=&1.93\times 10^{-31}\frac{c^2}{W_0^6}\,,
\end{eqnarray}
which can therefore vary in a large range of values consistent with all experimental bounds.
The rest of the masses do not present any significant change from the previous analysis in the dilaton case without the gauging.

\section{Conclusion}
\label{sect:conclusion}
In this work, we generalised the construction of new FI D-terms in $\mathcal{N}=1$ supergravity that do not require the gauging of R-symmetry and preserve invariance under ordinary K\"ahler transformations. Their bosonic part is just linear in the D-auxiliary field with a multiplicative factor which is an arbitrary function of the gravitino mass, expressed as a functional of the chiral multiplets. We then used these terms to construct new models of D-term inflation. Considering just a $U(1)$ and the inflaton multiplet with a no-scale K\"ahler potential and constant superpotential, we restricted to a simple form of the function associated to the new FI D-term which is a single positive power of the gravitino mass up to an additive constant. The later dominates the inflationary period by an asymptotically de Sitter regime, because the gravitino mass 
vanishes asymptotically in this region. The resulting models are consistent with observational CMB data and share common properties with the Starobinsky $R^2$ model~\cite{Starobinsky:1980te} on one hand and with the models of inflation by supersymmetry breaking on the other hand, where the inflaton is identified with the superpartner of the goldstino~\cite{Antoniadis:2017gjr, Antoniadis:2018cpq}. 
Moreover, they predict a variable range of primordial gravitational waves that can be within experimental reach. After the end of inflation, the inflaton rolls down to the minimum of the potential which has a tuneable (tiny) vacuum energy and supersymmetry is broken by a combination of F and D-term VEVs. 
An interesting open problem is whether there exists a microscopic origin of these new FI D-terms, for instance within string theory.

\acknowledgments
This work was supported in part by the Labex ``Institut Lagrange de Paris'', in part by the Swiss National Science Foundation, and in part by a CNRS PICS grant.

\appendix
\section{Multiplet calculus}
\label{sect:append_multiplet_calculus}
This appendix is based on \cite{multiplet_calculus}. A general complex scalar multiplet is given by
\begin{equation}
\mathcal{C}=(\mathcal{C}, \mathcal{Z}, \mathcal{H}, \mathcal{K}, \mathcal{B}_{\mu}, \Lambda,\mathcal{D}),
\end{equation}
where $\mathcal{C}$, $\mathcal{H}$, $\mathcal{K}$ and $\mathcal{D}$ are complex scalars, $\mathcal{Z}$ and $\Lambda$ are Dirac fermions, and $\mathcal{B}_{\mu}$ is a Lorentz vector. 
A chiral multiplet is obtained from a complex multiplet by imposing $P_R\mathcal{Z}=0$, $\mathcal{K}=0$, $\mathcal{B}_{\mu}=i\mathcal{D}_{\mu}\mathcal{C}$, $\Lambda=0$ and $\mathcal{D}=0$. Renaming $\mathcal{C}=Z$, it is written, in a seven-components notation, as
\begin{equation}
(Z, -i\sqrt{2}P_L\chi, -2F, 0, i\mathcal{D}_{\mu}Z, 0, 0),
\end{equation}
and similarly for its anti-chiral counterpart:
\begin{equation}
(\bar{Z}, i\sqrt{2}P_R\chi, 0, -2\bar{F}, -i\mathcal{D}_{\mu}\bar{Z}, 0, 0).
\end{equation}
The chiral and anti-chiral multiplets are also usually written in a three-components notation according to:
\begin{equation}
(Z, P_L\chi, F),\qquad(\bar{Z}, P_R\chi, \bar{F}).
\end{equation}

A real multiplet is obtained from a complex multiplet by imposing its lowest component $\mathcal{C}=C$ to be real. This implies $\mathcal{Z}=\zeta$ and $\Lambda=\lambda$ to be Majorana spinors, $\mathcal{B}_{\mu}=B_{\mu}$ and $\mathcal{D}=D$ to be real, while $\mathcal{K}=\bar{\mathcal{H}}$ is still complex. A real multiplet is thus written in a six-components notation according to
\begin{equation}
(C, \zeta, \mathcal{H}, B_{\mu}, \lambda, D).
\end{equation}   

Throughout this paper, the operation $[~]_F$ is defined as acting on a chiral multiplet $(Z,P_L\chi,F)$ of weights $(3,3)$ by:
\begin{equation}
[~]_F:(Z,P_L\chi,F)\rightarrow[Z]_F\equiv e~\left[F+\frac{1}{\sqrt 2}\bar\psi_{\mu}\gamma^{\mu}P_L\chi+\frac{1}{2}Z\bar\psi_{\mu}\gamma^{\mu\nu}P_R\psi_{\nu}\right]+h.c.
\end{equation}
The operation $[~]_D$ is defined as acting on a real multiplet $(C, \zeta, \mathcal{H}, B_{\mu}, \lambda, D)$ of weights $(2,0)$ by:
\begin{multline}\label{eq:D_operation}
[~]_D:(C, \zeta, \mathcal{H}, B_{\mu}, \lambda, D)\rightarrow[C]_D\equiv \frac{e}{2}\left[D-\frac{1}{2}\bar\psi_{\mu}\gamma^{\mu}i\gamma_*\lambda-\frac{1}{3}CR(\omega)+\frac{1}{6}\left(C\bar\psi_{\mu}\gamma^{\mu\rho\sigma}-i\bar\zeta\gamma^{\rho\sigma}\gamma_*\right)R'_{\rho\sigma}(Q)\right.\\\left.+\frac{1}{4}\epsilon^{abcd}\bar\psi_a\gamma_b\psi_c\left(B_d-\frac{1}{2}\bar\psi_d\zeta\right)\right],
\end{multline}
where $R(\omega)$ and $R'_{\rho\sigma}(Q)$ are the graviton and gravitino curvatures. Both operations are used to build superconformal invariant actions from chiral and real multiplets, respectively, according to $S_F=\int d^4x e~[Z]_F$ and $S_D=\int d^4x e~[C]_D$.\\

Given a set of complex multiplets $\mathcal{C}^i=(\mathcal{C}^i, \mathcal{Z}^i, \mathcal{H}^i, \mathcal{K}^i, \mathcal{B}_{\mu}^i, \Lambda^i,\mathcal{D}^i)$, $i=1...n$, one can build another complex multiplet $\mathcal{C}=(\mathcal{C}, \mathcal{Z}, \mathcal{H}, \mathcal{K}, \mathcal{B}_{\mu}, \Lambda,\mathcal{D})$ whose lowest component is given by an arbitrary function $f$ of the first components of $\mathcal{C}^i$'s: $\mathcal{C}=f(\mathcal{C}^i)$. The other components of $\mathcal{C}$ are then given by:
\begin{eqnarray}\label{eq:multiplet_composition_laws}
\mathcal{Z}&=&f_i\mathcal{Z}^i, \nonumber \\
\mathcal{H}&=&f_i\mathcal{H}^i-\frac{1}{2}f_{ij}\mathcal{\bar{Z}}^iP_L\mathcal{Z}^j, \nonumber \\
\mathcal{K}&=&f_i\mathcal{K}^i-\frac{1}{2}f_{ij}\mathcal{\bar{Z}}^iP_R\mathcal{Z}^j, \nonumber \\
\mathcal{B}_{\mu}&=&f_i\mathcal{B}_{\mu}^i+\frac{1}{2}if_{ij}\mathcal{\bar{Z}}^iP_L\gamma_{\mu}\mathcal{Z}^j, \\
\Lambda&=&f_i\Lambda^i+\frac{1}{2}f_{ij}\left(i\gamma_{*}\slashed{\mathcal{B}}^i+P_L\mathcal{K}^i+P_R\mathcal{H}^i-\slashed{\mathcal D}\mathcal{C}^i \right)\mathcal{Z}^j-\frac{1}{4}f_{ijk}\mathcal{Z}^i \mathcal{\bar{Z}}^j \mathcal{Z}^k, \nonumber \\
\mathcal{D}&=&f_i\mathcal{D}^i+\frac{1}{2}f_{ij}\left(\mathcal{K}^i\mathcal{H}^j-\mathcal{B}^i\cdot\mathcal{B}^j-\mathcal{D}\mathcal{C}^i\cdot\mathcal{D}\mathcal{C}^j-2\bar{\Lambda}^i\mathcal{Z}^j-\mathcal{\bar{Z}}^i\slashed{\mathcal{D}}\mathcal{Z}^j \right), \nonumber \\ &-&\frac{1}{4}f_{ijk}\mathcal{\bar{Z}}^i\left(i\gamma_{*}\slashed{\mathcal{B}}^j+P_L\mathcal{K}^j+P_R\mathcal{H}^j \right)\mathcal{Z}^k +\frac{1}{8}f_{ijkl}\mathcal{\bar{Z}}^iP_L\mathcal{Z}^j\mathcal{\bar{Z}}^k P_R\mathcal{Z}^l,\nonumber
\end{eqnarray}
with $f_i\equiv\frac{\partial f}{\partial \mathcal{C}^i}$ and so on for higher order derivatives. The bar on spinors are the Majorana conjugate defined by $\bar\psi=\psi^TC$, with $C$ the charge conjugation matrix satisfying $\gamma_{\mu}^T=-C\gamma_{\mu}C^{-1}$.

\section{Fermion masses}
\label{sect:append_fermion_masses}
In this appendix, we compute the fermion masses in the no-scale models presented in Section \ref{sect:inflation}. We recall the full Lagrangian considered in this paper:
\begin{equation}\label{eq:full_lagrangian_append}
\mathcal{L}=-3\left[S_0\bar{S}_0 e^{-\frac{\mathcal{K}(X,\bar X)}{3}}\right]_D+\left[S_0^3W(X)\right]_F-\frac{1}{4}\left[\bar{\lambda}P_L\lambda\right]_F+\mathcal{L}_{FI}^{(\alpha)}+\mathcal{L}_{FI}^{(-2/3)},
\end{equation}
with the new FI-terms given by:
\begin{equation}\label{eq:FI_lagrangian_3_append}
\mathcal{L}_{FI}^{(\alpha)}=-\xi\left[(S_0\bar{S}_0)^{-1}e^{(\frac{1}{3}+\alpha)K}W^{\alpha}\bar{W}^{\alpha}\frac{(\bar{\lambda}P_L\lambda)(\bar{\lambda} P_R\lambda)}{T(\bar{\mathcal{W}^2})\bar{T}(\mathcal{W}^2)}(V)_D \right]_D.
\end{equation}
We write the fermion mass terms as $\mathcal{L}_m=\mathcal{L}_m^{(0)}+\mathcal{L}_m^{FI}$, with $\mathcal{L}_m^{(0)}$ arising from the usual matter-coupled $\mathcal{N}=1$ supergravity Lagrangian, namely the three first terms of \eqref{eq:full_lagrangian_append}, and $\mathcal{L}_m^{FI}$ arising from the FI Lagrangians $\mathcal{L}_{FI}^{(\alpha)}+\mathcal{L}_{FI}^{(-2/3)}$. $\mathcal{L}_m^{(0)}$ reads:
\begin{equation}
e^{-1}\mathcal{L}_{\text{m}}^{(0)}=\frac{1}{2}m_{3/2}\bar{\psi}_{\mu}P_R\gamma^{\mu\nu}\psi_{\nu}+\bar{\psi}_{\mu}\gamma^{\mu}\left(\frac{1}{\sqrt{2}}\Omega^{\alpha}e^{K/2}\nabla_{\alpha}W+\frac{i}{2}\mathcal{P}_AP_L\lambda^A\right)-\frac{1}{2}m_{\alpha\beta}^{(0)}\bar{\Omega}^{\alpha}\Omega^{\beta}-m_{\alpha A}^{(0)}\bar{\Omega}^{\alpha}\lambda^A-\frac{1}{2}m_{AB}^{(0)}\bar{\lambda}^AP_L\lambda^B+h.c.,
\end{equation}
where $\psi_{\mu}$ denotes the gravitino, $\Omega^{\alpha}$ the chiral fermions, and $\lambda^A$ the gauginos. The various masses are given by \cite{Van_Proeyen}:
\begin{eqnarray}
m_{3/2}&=&e^{K/2}W,\\
m_{\alpha\beta}^{(0)}&=&e^{K/2}\nabla_{\alpha}\nabla_{\beta}W\equiv e^{K/2}\left(\partial_{\alpha}+\partial_{\alpha}K\right)\nabla_{\beta}W-e^{K/2}\Gamma_{\alpha\beta}^{\gamma}\nabla_{\gamma}W,\\
m_{\alpha A}^{(0)}&=&i\sqrt{2}\left(\partial_{\alpha}\mathcal{P}_A-\frac{1}{4}f_{AB,\alpha}(\text{Re}f)^{-1~BC}\mathcal{P}_C\right)=m_{A\alpha}^{(0)},\\
m_{AB}^{(0)}&=&-\frac{1}{2}e^{K/2}f_{AB,\alpha}g^{\alpha\bar{\beta}}\bar{\nabla}_{\bar{\beta}}\bar{W}.
\end{eqnarray}

In the no-scale models with K\"ahler potential $K=-p\ln(X+\bar X)$ studied in this paper, we consider only one chiral matter multiplet and one gauge multiplet, therefore the index $\alpha$ and $A$ take only one value. The Christoffel symbols are given by $\Gamma_{\alpha\beta}^{\gamma}=g^{\gamma\bar{\delta}}\partial_{\alpha}g_{\beta\bar{\delta}}$, which reduce to only one non-vanishing component $\Gamma_{XX}^X=-\frac{2}{X+\bar{X}}$. The moment map $\mathcal{P}$ is defined by $\mathcal{P}=i(k^{\alpha}\partial_{\alpha}K-r)$, where $k^{\alpha}$ is the Killing vector associated to the gauged symmetry and $r$ is the corresponding FI constant. In Section~\ref{sect:inflation} we considered $r=0$. When the chiral multiplet becomes charged under the gauged shift symmetry, the associated constant Killing vector is $k^X=ic$. Focussing on the ungauged case considered in Section~\ref{sect:inflation}, we thus have $\mathcal{P}=0$. Finally, the gauge kinetic function $f_{AB}$ being constant, we end up with:
\begin{equation}\label{eq:fermions_mass_sugra_Lagrangian}
m_{3/2}=e^{K/2}W,~~~
m_{\Omega\Omega}^{(0)}=m_{3/2}\frac{p(p-1)}{(X+\bar X)^{2}},~~~
m_{\Omega\lambda}^{(0)}=0,~~~
m_{\lambda\lambda}^{(0)}=0.
\end{equation}

We now turn to the fermion mass contributions arising from the FI Lagrangians.  
Keeping only quadratic terms in fermions containing no derivatives, the FI Lagrangian \eqref{eq:FI_lagrangian_3_append} can be written as
\begin{equation}
\mathcal{L}_{FI}^{(\alpha)}=-\xi\frac{D(W\bar{W})^{\alpha}}{T(\bar{\mathcal{W}^2})\bar{T}(\mathcal{W}^2)}\left[\hat{\mathcal{R}}^{(\alpha)}\right]_D,
\end{equation}
with $\hat{\mathcal{R}}^{(\alpha)}$ the real multiplet defined by 
\begin{equation}
\hat{\mathcal{R}}^{(\alpha)}\equiv e^{(\frac{1}{3}+\alpha)K(X,\bar X)}(S_0^{-1}\bar{\lambda}P_L\lambda)(\bar{S}_0^{-1}\bar{\lambda} P_R\lambda),
\end{equation}
which is a function of the chiral multiplets $S_0^{-1}\bar{\lambda}P_L\lambda$, $X$, and their anti-chiral counterparts. In the seven-components notation, they are given by:
\begin{eqnarray}
S_0^{-1}\bar{\lambda}P_L\lambda&=&(s_0^{-1}\bar{\lambda}P_L\lambda,~2s_0^{-1}DP_L\lambda,~2s_0^{-2}F_0\bar{\lambda}P_L\lambda+2s_0^{-1}D^2,~0,~0,~0,~0),\\
X&=&(X,~-i\sqrt{2}P_L\Omega,~-2F,~0,~0,~0,~0),\\
\bar{S}_0^{-1}\bar{\lambda} P_R\lambda&=&(\bar{s}_0^{-1}\bar{\lambda} P_R\lambda,~2\bar{s}_0^{-1}DP_R\lambda,~0,~2\bar{s}_0^{-2}\bar{F}_0\bar{\lambda} P_R\lambda+2\bar{s}_0^{-1}D^2,~0,~0,~0),\\
\bar{X}&=&(\bar{X},~i\sqrt{2}P_R\Omega,~0,~-2\bar{F},~0,~0,~0),
\end{eqnarray}
Writing $\hat{\mathcal{R}}^{(\alpha)}\equiv(0,0,0,(\hat{\mathcal{R}}^{(\alpha)})_{\mathcal{B}_{\mu}},(\hat{\mathcal{R}}^{(\alpha)})_{\lambda},(\hat{\mathcal{R}}^{(\alpha)})_{D})$, its contribution to the fermion masses arises from:
\begin{equation}
\left[\hat{\mathcal{R}}^{(\alpha)}\right]_D=\frac{e}{2}\left((\hat{\mathcal{R}}^{(\alpha)})_{D}-\frac{i}{2}\bar{\psi}\cdot\gamma\gamma_*(\hat{\mathcal{R}}^{(\alpha)})_{\lambda}\right).
\end{equation}
The tensor calculus \eqref{eq:multiplet_composition_laws} gives:
\begin{eqnarray}
(\hat{\mathcal{R}}^{(\alpha)})_{D}&=&\left[i2\sqrt{2}\partial_Xe^{(\alpha+1/3)K}s_0^{-1}\bar{s}_0^{-1}D^3\bar{\lambda}P_L\Omega-2\partial_Xe^{(\alpha+1/3)K}s_0^{-1}\bar{s}_{0}^{-1}D^2F\bar{\lambda}P_L\lambda\right.\nonumber\\&+&\left.2e^{(\alpha+1/3)K}\bar{s}_0^{-1}s_0^{-2}D^2F_0\bar{\lambda}P_L\lambda\right]+h.c.+2e^{(\alpha+1/3)K}\bar{s}_0^{-1}s_0^{-1}D^4,\\
(\hat{\mathcal{R}}^{(\alpha)})_{\lambda}&=&2e^{(\alpha+\frac{1}{3})K}D^3s_0^{-1}\bar{s}_0^{-1}\lambda.
\end{eqnarray}
Combining this with
\begin{equation}
\frac{D(W\bar{W})^{\alpha}}{T(\bar{\mathcal{W}^2})\bar{T}(\mathcal{W}^2)}=(W\bar W)^{\alpha+2/3}D^{-3}s_0^2\bar{s}_0^{2}-2(W\bar W)^{\alpha+2/3}\left(s_0\bar{s}_0^2\frac{F_0}{D^5}\bar{\lambda}P_L\lambda+h.c.\right),
\end{equation}
one obtains: 

(i) The gravitino-gaugino mixing:
\begin{equation}
e^{-1}\mathcal{L}_{FI}^{(\alpha)}\supset\frac{i}{2}\bar{\psi}_{\mu}\gamma^{\mu}\xi e^{(\alpha+2/3)\mathcal{G}}\gamma_*\lambda.
\end{equation}
Considering $\mathcal{L}^{(0)}+\mathcal{L}_{FI}^{(\alpha)}+\mathcal{L}_{FI}^{(-2/3)}$, we get the following gravitino/spin-$1/2$ mixing Lagrangian:
\begin{equation}\label{eq:mixing_lagrangian}
e^{-1}\mathcal{L}_{\text{mix}}=\bar{\psi}_{\mu}\gamma^{\mu}\left(\frac{1}{\sqrt{2}}\Omega e^{K/2}\nabla_{X}W+\frac{i}{2}(\xi_1e^{(\alpha+2/3)\mathcal{G}}+\xi_2)P_L\lambda\right)+h.c.,
\end{equation}
from which we identify the Goldstino as the linear combination:
\begin{equation}\label{eq:Goldstino}
P_L\nu=-\frac{1}{\sqrt{2}}\Omega e^{K/2}\nabla_{X}W-\frac{i}{2}(\xi_1e^{(\alpha+2/3)\mathcal{G}}+\xi_2)P_L\lambda.
\end{equation}

(ii) The fermion mass terms:
\begin{eqnarray}\label{eq:fermions_mass_FI_Lagrangian_1}
e^{-1}\mathcal{L}_{FI}^{(\alpha)}&\supset&-i\sqrt{2}\xi(\alpha+\frac{1}{3})\partial_XKe^{(\alpha+1/3)\mathcal{G}}(W\bar W)^{1/3}s_0\bar{s}_0\bar{\lambda}P_L\Omega+\xi(\alpha+\frac{1}{3})\partial_XKe^{(\alpha+1/3)\mathcal{G}}(W\bar W)^{1/3}s_0\bar{s}_0\frac{F}{D}\bar{\lambda}P_L\lambda\nonumber\\&+&\xi e^{(\alpha+\frac{1}{3})\mathcal{G}}(W\bar W)^{1/3}s_0\bar{s}_0\frac{s_0^{-1}F_0}{D}\bar{\lambda}P_L\lambda+h.c.
\end{eqnarray}
Considering from now on $\mathcal{L}_{FI}^{(\alpha)}+\mathcal{L}_{FI}^{(-2/3)}$, we define for simplicity of the expressions the following quantities:
\begin{eqnarray}
D_{\text{bos}}&\equiv&\left(\xi_1e^{(\alpha_1+\frac{1}{3})\mathcal{G}}+\xi_2e^{-\frac{1}{3}\mathcal{G}}\right)(W\bar W)^{1/3}s_0\bar{s}_0\nonumber\\
&=&\xi_1e^{(\alpha_1+\frac{2}{3})\mathcal{G}}+\xi_2,\\
\gamma&\equiv&\partial_XK\left(\xi_1(\alpha_1+\frac{1}{3})e^{(\alpha_1+\frac{1}{3})\mathcal{G}}-\frac{\xi_2}{3}e^{-\frac{1}{3}\mathcal{G}}\right)(W\bar W)^{1/3}s_0\bar{s}_0\nonumber\\
&=&\partial_XK\left(\xi_1(\alpha_1+\frac{1}{3})e^{(\alpha_1+\frac{2}{3})\mathcal{G}}-\frac{\xi_2}{3}\right).
\end{eqnarray}
One can already read from \eqref{eq:fermions_mass_FI_Lagrangian_1} the gaugino/chiral fermion mixing mass term:
\begin{equation}\label{eq:gaugino_chiral_mass}
m_{\Omega\lambda}^{FI}=i\sqrt{2}\gamma.
\end{equation} 
In order to find the gaugino mass term $m_{\lambda\lambda}^{FI}$, we have to eliminate the auxiliary fields $D$, $F$ and $F_0$ using their equations of motion. The part of the total Lagrangian containing the auxiliary field $D$, up to quadratic order in fermions, is:
\begin{equation}\label{eq:L_with_D}
e^{-1}\mathcal{L}\supset\frac{1}{2}D^2-D_{\text{bos}}D+\left(D_{\text{bos}}\frac{s_0^{-1}F_0}{D}\bar{\lambda}P_L\lambda+\gamma\frac{F}{D}\bar{\lambda}P_L\lambda+h.c.\right),   
\end{equation}
so that the equation of motion for $D$ reads $D^3-D_{\text{bos}}D^2-\left[(D_{\text{bos}}s_0^{-1}F_0+\gamma F)\bar{\lambda}P_L\lambda+h.c.\right]=0$. Solving it analytically and expanding the solution up to quadratic order in fermions, we find 
\begin{equation}\label{eq:D}
D=D_{\text{bos}}+\left(\frac{D_{\text{bos}}s_0^{-1}F_0+\gamma F}{D_{\text{bos}}^2}\bar{\lambda}P_L\lambda+h.c.\right)+\text{higher order in fermions}.
\end{equation}
Replacing \eqref{eq:D} in \eqref{eq:L_with_D}, we find the following quadratic contribution in fermions:
\begin{equation}
e^{-1}\mathcal{L}\supset \frac{D_{\text{bos}}s_0^{-1}F_0+\gamma F}{D_{\text{bos}}}\bar{\lambda}P_L\lambda+h.c.
\end{equation}

We now eliminate the auxiliary fields $F_0$ and $F$, associated to the compensator and $X$ chiral multiplet, respectively. The part of the total Lagrangian containing the auxiliary fields $F_0$, $F$, up to the quadratic order in fermions, reads:
\begin{eqnarray}
e^{-1}\mathcal{L}\supset &-&3e^{-K/3}F_0\bar{F}_0+3e^{K/3}WF_0+3e^{K/3}\bar{W}\bar{F}_0+\frac{1}{9}g_{X\bar X}F\bar{F}+\frac{1}{3}e^{K/2}\nabla_X WF+\frac{1}{3}e^{K/2}\bar{\nabla}_{\bar X}\bar{W}\bar{F}\nonumber\\
&+&\frac{1}{D_{\text{bos}}}(D_{\text{bos}}e^{-K/6}F_0+\gamma F)\bar{\lambda}P_L\lambda+\frac{1}{D_{\text{bos}}}(D_{\text{bos}}e^{-K/6}\bar{F}_0+\gamma \bar{F})\bar{\lambda} P_R\lambda,
\end{eqnarray}
which yields, after elimination of $F_0$ and $F$:
\begin{equation}
e^{-1}\mathcal{L}\supset e^K\left(3W\bar W-\nabla_X Wg^{X\bar X}\bar{\nabla}_{\bar X}\bar{W}\right)+\frac{e^{K/2}}{D_{\text{bos}}}\left(D_{\text{bos}}\bar{W}\bar{\lambda}P_L\lambda-3g^{X\bar X}\bar{\nabla}_{\bar X}\bar{W}\gamma\bar{\lambda}P_L\lambda+h.c.\right).
\end{equation}
The first two terms correspond to the usual $F$-contribution to the scalar potential, while the last two terms give the contribution to the gaugino mass $m_{\lambda\lambda}^{FI}$ from the FI-terms. For a constant superpotential, it reads:
\begin{equation}\label{eq:gaugino_gaugino_mass}
m_{\lambda\lambda}^{FI}=-2\frac{m_{3/2}}{D_{\text{bos}}}\left[D_{\text{bos}}+3(X+\bar X)\gamma\right].
\end{equation}
At the minimum of the potential where $\partial_X \mathcal{V}=0$ and $\mathcal{V}=0$, $m_{\Omega\Lambda}^{FI}$ and $m_{\Lambda\Lambda}^{FI}$ given in Eqs.~\eqref{eq:gaugino_chiral_mass} and \eqref{eq:gaugino_gaugino_mass} simplify, and the entries of the fermion mass matrix can be written as: 
\begin{eqnarray}\label{eq:fermion_masses_min}
\label{eq:ferm_mass_chi_chi}
m_{\Omega\Omega}&=&m_{\Omega\Omega}^{(0)}=m_{3/2}\frac{p(p-1)}{(X+\bar X)^{2}},\nonumber\\
\label{eq:ferm_mass_chi_gaug}
m_{\Omega\lambda}&=&m_{\Omega\lambda}^{FI}=-\frac{i\sqrt{2}}{6}\frac{pD_{\text{bos}}}{X+\bar X},\nonumber\\
\label{eq:ferm_mass_gaug_gaug}
m_{\lambda\lambda}&=&m_{\lambda\lambda}^{FI}=-2m_{3/2}\left(1-\frac{p}{2}\right).
\end{eqnarray}

In order to study the spin-1/2 fermions mass matrix, we have to get rid of the gravitino-Goldstino mixing  \eqref{eq:mixing_lagrangian}. This can be done by carrying out a supersymmetry transformation, bringing the gravitino $\psi_{\mu}$ into the physical, massive, one $\Psi_{\mu}$ through \cite{Van_Proeyen}:
\begin{equation}\label{eq:gravitino_transformation}
P_L\psi_{\mu}\rightarrow P_L\Psi_{\mu}=P_L\psi_{\mu}-\frac{2}{3m_{3/2}^2}\partial_{\mu}P_L\nu-\frac{1}{3m_{3/2}}\gamma_{\mu}P_R\nu.
\end{equation}
The mixing term between the gravitino and the Goldstino then vanishes, and $\Psi_{\mu}$ is the massive gravitino in Minkowski space with physical mass $m_{3/2}$. In addition, the transformation \eqref{eq:gravitino_transformation} brings new contributions to the spin-1/2 fermion mass terms. Writing the Goldstino $P_L\nu$ as a linear combination of the gaugino $\lambda$ and the chiral fermion $\Omega$, namely $P_L\nu=A\Omega+BP_L\lambda$ where $A$ and $B$ are given in this model by Eq.~\eqref{eq:Goldstino}, these new contributions read: 
\begin{eqnarray}\label{eq:fermions_shifts}
m_{\Omega\Omega}^{(\nu)}&=&-\frac{4}{3m_{3/2}^2}A^2,\nonumber\\
m_{\Omega\lambda}^{(\nu)}&=&-\frac{4}{3m_{3/2}^2}AB,\nonumber\\
m_{\lambda\lambda}^{(\nu)}&=&-\frac{4}{3m_{3/2}^2}B^2.
\end{eqnarray}
The most general structure of a spin-$1/2$ mass term $m^{(g)}$ is then given by $m^{(g)}\equiv m^{(0)}+m^{FI}+m^{(\nu)}$, with $m^{(0)}$ the contribution from the original Lagrangian $-3\left[S_0\bar{S}_0 e^{-\frac{\mathcal{K}(X,\bar X)}{3}}\right]_D+\left[S_0^3W(X)\right]_F-\frac{1}{4}\left[\bar{\lambda}P_L\lambda\right]_F$, $m^{FI}$ the contribution from the new FI terms $\mathcal{L}_{FI}^{(\alpha)}+\mathcal{L}_{FI}^{(-2/3)}$, and $m^{(\nu)}$ the shifts~\eqref{eq:fermions_shifts} upon elimination of the gravitino-Goldstino mixing. From Eqs.~\eqref{eq:fermions_mass_sugra_Lagrangian}, \eqref{eq:fermion_masses_min} and \eqref{eq:fermions_shifts}, we deduce the fermion mass matrix at the minimum of the potential:
\begin{equation}
M=
\begin{pmatrix} 
m_{\Omega\Omega}^{(g)} & m_{\Omega\lambda}^{(g)} \\ 
m_{\Omega\lambda}^{(g)} & m_{\lambda\lambda}^{(g)} 
\end{pmatrix}
=
\begin{pmatrix} 
\frac{p-3}{3}m_{3/2} & \frac{i\sqrt{2}}{6}\sqrt{p}D_{\text{bos}} \\ 
\frac{i\sqrt{2}}{6}\sqrt{p}D_{\text{bos}} & \frac{p}{3}m_{3/2} 
\end{pmatrix}
\end{equation}
A normalisation factor for $m_{\Omega\Omega}^{(g)}$ and $m_{\Omega\lambda}^{(g)}$ has been introduced due to the non-canonical kinetic term of the chiral fermion, while the gaugino already has canonical kinetic term since the gauge kinetic function $f$ has been set to one. Using $D_{\text{bos}}^2=-2(p-3)m_{3/2}^2$ at the minimum of the potential, one immediately sees that the determinant of $M$ vanishes, while its non-zero eigenvalue $m_f$, corresponding to the mass of the physical fermion, is given by:
\begin{equation}\label{eq:physical_fermion_mass}
m_f^{2}=m_{3/2}^2\left(\frac{4}{9}p^2-\frac{4}{3}p+1\right), \forall p\neq 3\,,
\end{equation}
where we excluded the value $p=3$ for which, in the case of a constant superpotential, the D-term vanishes in the minimum, making the new FI-term singular, and a different superpotential is used in Section \ref{sect:volume_sector}.


\begin{thebibliography}{10}

\bibitem{Van_Proeyen}
D.~Z.~Freedman and A.~Van Proeyen, "Supergravity", Cambridge, UK: CambridgeUni.~Pr.~(2012) 607p

\bibitem{Antoniadis:2014oya}
  I.~Antoniadis, E.~Dudas, S.~Ferrara and A.~Sagnotti,
  ``The Volkov-Akulov-Starobinsky supergravity,''
  Phys.\ Lett.\ B {\bf 733} (2014) 32
  [arXiv:1403.3269 [hep-th]].

\bibitem{Freedman_model_1}
D.~Z.~Freedman, "Supergravity with Axial Gauge Invariance", Phys.~Rev.~D15 (1977) 1173.

\bibitem{Freedman_model_2}
R.~Barbieri, S.~Ferrara, D.~V.~Nanopoulos and K.~S.~Stelle, "Supergravity, R Invariance and Spontaneous Supersymmetry Breaking", Phys.~Lett.~113B (1982) 219.

\bibitem{new_FI_1}
N.~Cribiori, F.~Farakos, M.~Tournoy, A.~Van Proeyen, "Fayet-Iliopoulos terms in supergravity without gauged R-symmetry", J.~High Energ.~Phys.~(2018) 2018: 32
 [arXiv:1712.08601 [hep-th]].

\bibitem{Antoniadis:2018cpq}
  I.~Antoniadis, A.~Chatrabhuti, H.~Isono and R.~Knoops,
  ``Fayet-Iliopoulos terms in supergravity and D-term inflation,''
  Eur.\ Phys.\ J.\ C {\bf 78} (2018) no.5,  366
  [arXiv:1803.03817 [hep-th]].

\bibitem{new_FI_2}
I.~Antoniadis, A.~Chatrabhuti, H.~Isono, R.~Knoops, "The cosmological constant in supergravity", Eur.~Phys.~J.~C (2018) 78: 718
[arXiv:1805.00852 [hep-th]].
\bibitem{Antoniadis:2019hbu}
  I.~Antoniadis, J.~P.~Derendinger, F.~Farakos and G.~Tartaglino-Mazzucchelli,
  ``New Fayet-Iliopoulos terms in $ \mathcal{N}=2 $ supergravity,''
  JHEP {\bf 1907} (2019) 061
  [arXiv:1905.09125 [hep-th]].

\bibitem{Cremmer:1983bf}
  E.~Cremmer, S.~Ferrara, C.~Kounnas and D.~V.~Nanopoulos,
  ``Naturally Vanishing Cosmological Constant in $ \mathcal{N}=1 $ Supergravity,''
  Phys.\ Lett.\  {\bf 133B} (1983) 61;
  J.~R.~Ellis, C.~Kounnas and D.~V.~Nanopoulos,
  ``Phenomenological SU(1,1) Supergravity,''
  Nucl.\ Phys.\ B {\bf 241} (1984) 406.
  
\bibitem{Antoniadis:2017gjr}
  I.~Antoniadis, A.~Chatrabhuti, H.~Isono and R.~Knoops,
  ``Inflation from Supersymmetry Breaking,''
  Eur.\ Phys.\ J.\ C {\bf 77} (2017) no.11,  724
  [arXiv:1706.04133 [hep-th]].
  
\bibitem{AlvarezGaume:2010rt}
  L.~Alvarez-Gaume, C.~Gomez and R.~Jimenez,
  ``Minimal Inflation,''
  Phys.\ Lett.\ B {\bf 690} (2010) 68
  [arXiv:1001.0010 [hep-th]];
  ``A Minimal Inflation Scenario,''
  JCAP {\bf 1103} (2011) 027
  [arXiv:1101.4948 [hep-th]].

\bibitem{mass_de_sitter_space_1}
G.~B\"orner and H.~P.~D\"urr, {\it Classical and quantum fields in de Sitter space}, Nuovo Cimento A (1965-1970) (1969) 64:669

\bibitem{mass_de_sitter_space_2}
A.~B\"ohm, {\it Dynamical group and mass spectrum}, Phys.~Rev.~145 (1965) 1212

\bibitem{GS}
  M.~Dine, N.~Seiberg and E.~Witten,
  ``Fayet-Iliopoulos Terms in String Theory,''
  Nucl.\ Phys.\ B {\bf 289} (1987) 589.

\bibitem{Starobinsky:1980te}
  A.~A.~Starobinsky,``A New Type of Isotropic Cosmological Models Without Singularity,''  
  Phys.\ Lett.\  {\bf 91B} (1980) 99.

\bibitem{multiplet_calculus}
S.~Ferrara, R.~Kallosh, A.~Van Proeyen and T.~Wrase, {\it Linear Versus Non-linear Supersymmetry, in General}, JHEP {\bf 1604} (2016) 065
[arXiv:1603.02653 [hep-th]].

\end{thebibliography}
\end{document}